\documentclass[a4paper,twocolumn,superscriptaddress,accepted=2017-12-06]{quantumarticle}
\usepackage[numbers,sort&compress]{natbib}
\usepackage{amsmath}
\usepackage{graphicx}
\usepackage[utf8]{inputenc}
\usepackage[english]{babel}
\usepackage[T1]{fontenc}
\usepackage[colorlinks]{hyperref}

\pdfoutput=1

\newcommand{\braket}[1]{\left<#1\right>}

\newcommand{\eps}{\epsilon}
\newcommand{\Op}[1]{\hat{#1}}
\newcommand{\oa}{\Op{a}}
\newcommand{\oU}{\Op{U}}
\newcommand{\oH}{\Op{H}}
\newcommand{\oL}{\Op{L}}

\begin{document}

\title{Quantum and classical dynamics of a three-mode absorption refrigerator}%

\author{Stefan Nimmrichter}
 \email{cqtsn@nus.edu.sg}
 \orcid{0000-0001-9566-3824}
 \affiliation{Centre for Quantum Technologies, National University of Singapore, 3 Science Drive 2, Singapore 117543.}
\author{Jibo Dai}
 \affiliation{Centre for Quantum Technologies, National University of Singapore, 3 Science Drive 2, Singapore 117543.}
 \altaffiliation{Now at Data Storage Institute, A*STAR}
 \author{Alexandre Roulet}
 \affiliation{Centre for Quantum Technologies, National University of Singapore, 3 Science Drive 2, Singapore 117543.}
 \affiliation{Now at Department of Physics, University of Basel, Klingelbergstrasse 82, CH-4056 Basel, Switzerland}
\author{Valerio Scarani}
\orcid{0000-0001-5594-5616}
 \affiliation{Centre for Quantum Technologies, National University of Singapore, 3 Science Drive 2, Singapore 117543.}
 \affiliation{Department of Physics, National University of Singapore, 2 Science Drive 3, Singapore 117542.}


\begin{abstract}
We study the quantum and classical evolution of a system of three harmonic modes interacting via a trilinear Hamiltonian. With the modes prepared in thermal states of different temperatures, this model describes the working principle of an absorption refrigerator that transfers energy from a cold to a hot environment at the expense of free energy provided by a high-temperature work reservoir. 
Inspired by a recent experimental realization with trapped ions, we elucidate key features of the coupling Hamiltonian that are relevant for the refrigerator performance. The coherent system dynamics exhibits rapid effective equilibration of the mode energies and correlations, as well as a transient enhancement of the cooling performance at short times. We find that these features can be fully reproduced in a classical framework.
\end{abstract}

\maketitle

\section{Introduction}
Historically, thermodynamics started with the goal of trying to understand how to convert heat into useful mechanical motion. For that purpose, steam engines have been developed which revolutionized our lives. Other useful thermal machines, such as refrigerators and heat pumps, followed. Naturally, these thermal machines are all large macroscopic entities that one uses classical laws to describe them and study their performances.

Recently, as our ability to control small quantum systems progresses, it has now become not only interesting, but also crucial to study what happens when these machines become microscopic where quantum features become important \cite{Horodecki+Oppenheim:13, Skrzypczyk+2:14}. More specifically, there has been significant interest in determining the extent to which quantum effects may help surpass classical limits such as the Carnot efficiency \cite{Scully2003, Dillenschneider2009, Rossnagel+4:14, Correa+3:14,Strasberg2017,Niedenzu2017}, 
and in how to export classical notions such as work to the quantum domain \cite{Allahverdyan2004,Talkner2007,Campisi2011,Roncaglia+2:14, Talkner+Hanggi:16}. Various quantum thermal machines have been proposed and studied, with potential realizations in optomechanical setups, superconducting circuits, atom-cavity systems and trapped ions \cite{Geva1996,Venturelli+2:13, Kosloff+Levy:14, Rossnagel+4:14, Mari+2:15, Hofer+2:16, Joulain+3:16, Mitchison+4:16, Hofer+5:16, Karimi+Pekola:16, Roulet+3:16, Bissbort+5:16,Zhang2017,Reid2017,Hardal2017,Mu2017}.

In the context of refrigerators, two quantum approaches stand out: the smallest possible refrigerator consists of a qutrit \cite{Palao2001} or three interacting qubits \cite{Linden+2:10}, but a quantum absorption refrigerator can also be realised with three interacting harmonic modes \cite{Levy+Kosloff:12, Goold+4:16}. This latter interaction, which has been demonstrated with trapped ions \cite{Maslennikov+8:17}, is the object of this study. We focus on the unitary dynamics of the refrigerator itself and highlight phenomena like effective equilibration, as well as the challenge of identifying genuine quantum features. Elements on the process of refrigeration, i.e. the transfer of energy from a cold to a hot bath, mediated by the machine, are given in the Appendix.

Section~\ref{UnitaryDynamics} introduces the model Hamiltonian, the initial states that will be considered (independent thermal states of different temperatures) and the dynamical variables whose evolution is studied (occupation number of each mode). We discuss why we focus on the unitary dynamics. This dynamics is then solved and discussed in Section~\ref{secUnit}, first in general, then in the context of refrigeration. The following two sections are devoted to the two main observed features of the dynamics of the occupation numbers. In the short-term regime, there is a \textit{cooling enhancement} which would be absent if the interaction were incoherent (Section~\ref{SingleShot}). In the long-term regime, one observes \textit{effective equilibration} of the occupation numbers even if there is no dissipative dynamics (Section~\ref{QuantumEquilibration}). Finally, in Section~\ref{classicalModel} we study the classical model obtained by replacing the mode operators with conjugate variables in the Hamiltonian. We show that the dynamics of the occupation numbers exhibits the same features observed in the quantum formalism.

\section{Model}\label{UnitaryDynamics}
The model we study here is a system of three interacting harmonic oscillators, with the free Hamiltonian
\begin{equation}\label{freeHamiltonian}
\oH_0=\sum_{i=h,w,c}\hbar\omega_i\left(\oa_i^\dagger\oa_i+\frac{1}{2}\right),
\end{equation}
and the interaction Hamiltonian
\begin{equation}\label{Hinteraction}
\oH_1=\hbar g(\oa_h^\dagger \oa_w \oa_c+\oa_h\oa_w^\dagger\oa_c^\dagger).
\end{equation}
It is convenient to work in the interaction picture where
\begin{equation}\label{HinteractionPicture}
\oH_\mathrm{int}=\hbar g(\oa_h^\dagger \oa_w \oa_c \,e^{i\Delta t}+ \oa_h \oa_w^\dagger \oa_c^\dagger \,e^{-i\Delta t}),
\end{equation}
with $\Delta=\omega_h-\omega_w-\omega_c$ and $g$ the coupling constant. We will focus on the resonant case where $\Delta=0$ in what follows.

This Hamiltonian describes a wide range of physical processes: parametric amplification, frequency conversion, and second harmonic generation \cite{Walls+Barakat:70, Agrawal+Mehta:74, Gambini:77}. This work is based on considering the three interacting modes as an absorption refrigerator \cite{Levy+Kosloff:12, Goold+4:16, Maslennikov+8:17}.

The refrigeration process works as follows. Each of the three modes is in contact with a thermal bath: the cold bath at $T_c$, the hot bath at $T_h>T_c$, and the work mode at $T_w$. We shall correspondingly refer to $\oa_h$, $\oa_w$ and $\oa_c$ as the hot mode, work mode, and cold mode, respectively. The dynamical variables of interest are the occupation numbers $\bar{n}_i(t)=\mathrm{tr}\{\rho(t)\oa_i^\dagger \oa_i\}$ with $i=h,w$ or $c$. Absorption refrigeration occurs if the interaction can induce a stationary heat flow from the cold to the hot mode ($\dot{n}_c(t)<\dot{n}_c(0)$, $\dot{n}_h(t)>\dot{n}_h(0)$), the work mode providing a sufficient amount of free energy. 

A full description of the refrigeration process takes into account both the unitary interaction generated by the trilinear Hamiltonian and the dissipative dynamics due to the interaction of each mode with its bath. This is a numerically involved task, further complicated by the subtle issue of formulating a master equation for composite systems coupled to multiple baths that is consistent with the second law of thermodynamics \cite{Levy+Kosloff:14,Gonzalez2017,Hofer2017}. Yet, we are interested in the regime of weak coupling to the baths such that all the interesting aspects of the dynamics of refrigeration are captured by restricting the analysis to the unitary dynamics. This is confirmed in the Appendix, where we provide a comparison of the dynamics with and without the dissipative coupling. In the rest of this paper, we shall thus focus on the unitary dynamics, which has been experimentally implemented with trapped ions~\cite{Maslennikov+8:17}.

The initial state of the system consists in the three modes being prepared in uncorrelated thermal states, $\rho (t=0) = \rho_h^\mathrm{th}\otimes \rho_w^\mathrm{th}\otimes \rho_c^\mathrm{th}$ with
\begin{equation}\label{thermalState}
\rho_i^{\mathrm{th}}= \left[ 1 - \exp \left( - \frac{\hbar \omega_i}{k_B T_i} \right) \right] \exp \left( - \frac{\hbar \omega_i}{k_B T_i} \oa_i^\dagger \oa_i  \right).
\end{equation}
This corresponds to the situation in which each modes have been kept in contact with their respective baths for long time, before turning on the interaction. From here, focusing on the unitary dynamics implies that the modes are effectively decoupled from their baths during the evolution; this will capture the full refrigeration dynamics in the limit of slow thermalization rate.

Before proceeding, let us notice that some states of this family are stationary with respect to the unitary dynamics, and therefore also to the complete dynamics too. Indeed $[\rho_\text{st},\oH_{\mathrm{int}}]=0$ if
\begin{equation}\label{stationaryCond}
\frac{1}{\bar{n}_h(0)}+1=\left(\frac{1}{\bar{n}_w(0)}+1\right)\left(\frac{1}{\bar{n}_c(0)}+1\right).
\end{equation}

\section{Unitary dynamics}
\label{secUnit}

\subsection{Methods and general features}

We will mostly focus on temperatures $T_i$ that correspond to comparably small initial average occupation numbers, $\bar{n}_i(0)=\mathrm{tr}\{\rho(t=0)\oa_i^\dagger \oa_i\}$ with $i=h,w$ or $c$. 
For later comparison with the classical framework, we will plot the initial average energy of each mode in the diagrams, $\eps_i(0)=\hbar\omega_i(\bar{n}_i(0) + 1/2)$, instead of the occupation number.

At time $t=0$, the interaction Hamiltonian $\oH_\mathrm{int}$ is switched on and the system evolves unitarily according to $\rho(t)=\oU\rho_0 \oU^\dagger$ with $\oU=\mathrm{exp}(-i\oH_\mathrm{int} t)$. This coherently transfers populations between the three modes and changes the average energies $\epsilon_i(t)$. However, we note that even the closed system is not amenable to an exact analytical solution; previous studies focused on either short-time behavior, resorted to using products of coherent or Fock states as the initial state, or considered limiting cases of average occupation numbers much larger than one \cite{Walls+Barakat:70, Bonifacio+Preparata:70, Agrawal+Mehta:74, Gambini:77}.

For an efficient simulation of the system dynamics, we take advantage of the fact that the interaction Hamiltonian couples only Fock states of the form $|n_h,n_w,n_c \rangle = |n, N-n, M-n\rangle$ with fixed integers $N$ and $M$ and $0\leq n\leq \mathrm{min}\{N,M \}$. That is, $\oH_{\rm int}$ is block-diagonal with respect to finite-dimensional subspaces characterized by the two conserved quantities $N$ and $M$ and the dimension $d = \min\{N,M\} + 1$. The unitary evolution can then be efficiently computed by diagonalizing the Hamiltonian in each of the contributing subspaces, up to a cutoff for both $N$ and $M$ that truncates the originally infinite dimensional Hilbert space. For the simulations presented in this paper, the cutoff is chosen to ignore populations in the  initial density matrix smaller than $10^{-4}$.

\begin{figure}[t]		
\centering
\includegraphics[width=\columnwidth]{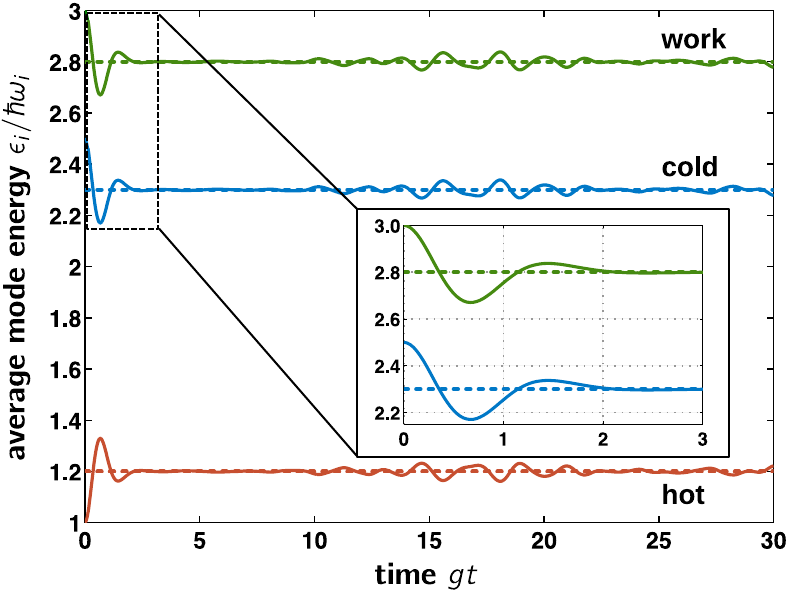}
\caption{Time evolution of the average energy (in units of $\hbar\omega_i$) of the hot, work and cold mode. The initial state is a product thermal state with average occupations 
$\bar{n}_h(0)=0.5$, $\bar{n}_w(0)=2.5$, and $\bar{n}_c(0)=2.0$. 
Time is measured in units of the inverse coupling strength $1/g$. The dashed lines show the values associated to the infinite time-averaged state \eqref{infTimeAve}. The initial transient oscillations are magnified in the inset.}
\label{fig:systemEvolution}
\end{figure}

Fig.~\ref{fig:systemEvolution} shows the typical time evolution of the modes' energy under the unitary dynamics for a thermal state of the form~\eqref{thermalState}. In particular, we have chosen initial average occupations of $\bar{n}_h(0)=0.5$, $\bar{n}_w(0)=2.5$, and $\bar{n}_c(0)=2.0$, so that the order of magnitude of the contributing $N$ and $M$ are $10^0$ or $10^1$. It is readily seen that $\bar{N}=\bar{n}_h(t)+\bar{n}_w(t)$, and $\bar{M}=\bar{n}_h(t)+\bar{n}_c(t)$ are conserved.

The curves in Fig.~\ref{fig:systemEvolution} exhibit two noteworthy features. Firstly, the evolution starts with a \textit{transient oscillatory stage}, where the largest fluctuation away from the initial value is found. We shall come back to this feature in Section~\ref{SingleShot}. Secondly, the system energies seem to approach some apparent equilibrium values (dashed lines), around which only small residual oscillations persist. One can obtain those values from the infinite time-averaged state \begin{equation}\label{infTimeAve}
\sigma\equiv\langle\rho(t)\rangle_\infty=\lim_{\tau\rightarrow\infty}\frac{1}{\tau}\int_0^\tau\text{d}t\,\rho(t).
\end{equation}
The observation that the expectation value of certain dynamical variables approach their long-time average, in spite of the fact that the system is not converging to a steady state because the dynamics is unitary, is a phenomenon known as effective equilibration \cite{Gogolin+Eisert:16}. We will get back to it in Section~\ref{QuantumEquilibration}. 

\begin{figure}[ht]
   \centering
   \includegraphics[width=\columnwidth]{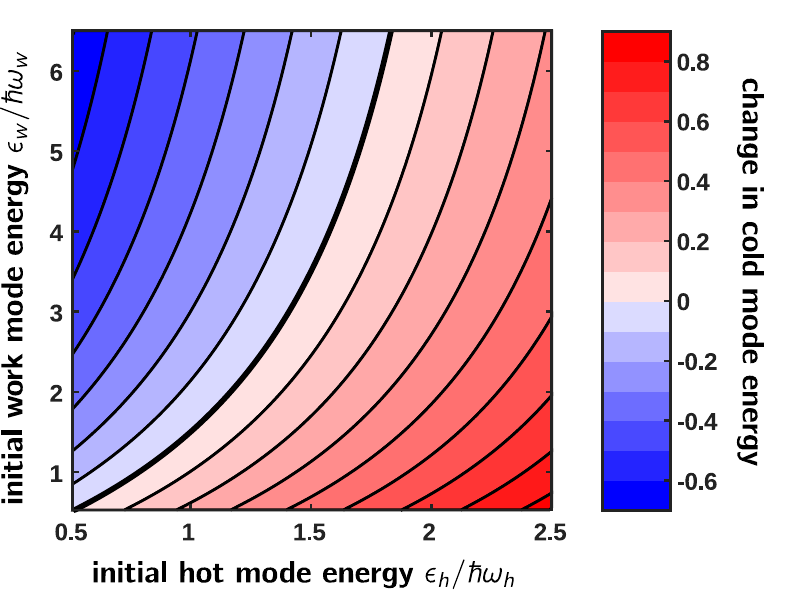}
        \caption{Change in the cold mode energy from the initial thermal average $\eps_c (0) = 2.5\hbar\omega_c$ to the infinite time-averaged value for varied initial energies of the hot and the work mode. All energies are given in units of the respective excitation quanta. The thick line indicates the stationary configurations of the interaction Hamiltonian, where no change occurs.}
    \label{fig:changeInCmode}
\end{figure}

Conservation of $N$ and $M$ implies that any increase in the average energy of the hot mode corresponds to an equal decrease in the work mode and cold mode energies. Hence it is sufficient to focus on just one of the modes. In Fig.~\ref{fig:changeInCmode}, we plot the change of the average cold mode energy (or occupation number) between the long-time equilibrium state $\sigma$ and the initial state for different initial energies of the work and the hot mode.
The initial occupation number in mode $c$ is fixed at $\bar{n}_c(0)=2.0$. One sees that, depending on the values of $\bar{n}_h(0)$ and $\bar{n}_w(0)$, the equilibrium energy of mode $c$ can either increase (red), decrease (blue), or stay unchanged (thick line). 

Let us remark again that, by preparing the initial system in a product of thermal states and then bringing them into interaction, we mimic the thermodynamic situation where these three subsystems each have thermalized with their own baths and are then allowed to exchange heat among each other. 
In Appendix~\ref{OpenSystem}, we explicitly compare this two-step treatment with a simultaneous interaction-dissipation model based on a heuristic master equation. The latter contains the interaction Hamiltonian and independent thermal dissipators for each mode. 
When the dissipation rates are small compared to the coupling rate $g$, we find that the time evolution of the system energies is similar to that of the purely unitary dynamics. Specifically, the initial transient features are approximately the same, and the only relevant difference is a slight deviation of the steady-state energies from the long-time averages predicted by the unitary model. Strong dissipation rates, on the other hand, would only thwart the three-mode interaction dynamics and freeze the system state close to the initial thermal state.

\subsection{Analogies and differences with refrigeration processes}\label{AbsFridge}

After presenting the general features of the closed system dynamics governed by the trilinear Hamiltonian \eqref{HinteractionPicture}, we now focus on the thermodynamic aspects. 
As already mentioned, the interaction is capable of driving a quantum absorption refrigerator, in close analogy to the earlier proposed three-qubit fridge \cite{Linden+2:10} with its interaction Hamiltonian $\oH_\mathrm{int}=g(|010\rangle\langle 101|+|101\rangle\langle 010|)$. The cooling performance of the latter has been extensively studied, and we will point out similarities and crucial differences for the present three-oscillator case.

\begin{figure}[t]
   \centering
        \includegraphics[width=\columnwidth]{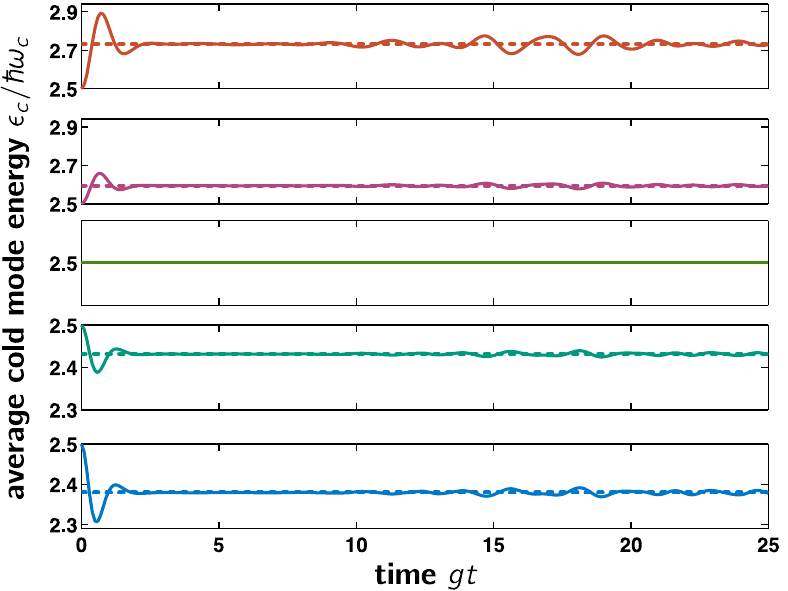}
        \caption{Time evolution of the average cold mode energy (in units of $\hbar\omega_c$) for five different initial conditions ranging from a net heating to a net cooling scenario (top to bottom). The initial hot and cold mode energies are fixed at $\eps_h=1.5 \hbar\omega_h$ and $\eps_c=2.5 \hbar\omega_c$, while the work mode energy starts from $\eps_w=1.5 \hbar\omega_w$ in the top panel and is increased in steps of $\hbar\omega_w$ to the bottom. Note that this still admits a lower temperature in the cold than in the hot mode, as $\omega_c < \omega_h$. The middle panel corresponds to a stationary configuration.}
    \label{fig:heating2cooling}
\end{figure}

As proposed in theory \cite{Levy+Kosloff:12} and backed by observations with normal modes of trapped ytterbium ions \cite{Maslennikov+8:17}, the average energy of the $c$-mode can decrease or increase, depending on the initial work mode value $\bar{n}_w(0)$. In this sense, we can speak of cooling or heating, as demonstrated in Fig.~\ref{fig:heating2cooling} for an exemplary choice of initial values, $\bar{n}_h(0)=1.0$, $\bar{n}_c(0)=2.0$, and $\bar{n}_w(0)$ varying from $1.0$ to $5.0$ in unit steps.
In fact, the contour plot in Fig.~\ref{fig:changeInCmode} can be viewed as a phase diagram, where the thick solid line separates the heating (red) from the cooling (blue) regime. For the latter, the initial occupation numbers must satisfy
\begin{equation}
\bar{n}_w (0)>\bar{n}_h (0) \frac{1+\bar{n}_c (0)}{\bar{n}_c (0) - \bar{n}_h (0)}. \label{eq:coolingRegime}
\end{equation}
Since the initial states are thermal, the inequality translates into a relation between the temperatures, $T_i=\hbar\omega_i / k_B\text{ln}(1+1/\bar{n}_i (0))$. Notice here that the refrigerator setting $T_c < T_h$ does not necessarily imply $\bar{n}_c(0)<\bar{n}_h(0)$ due to the different eigenfrequencies.

The cooling regime \eqref{eq:coolingRegime} is consistent with the general \emph{virtual qubit} framework developed recently \cite{Brunner+3:12}. In this picture, one thinks of the hot mode and work mode forming a set of virtual qubits by the levels $|n_h,n_w\rangle$ and $|n_h-1,n_w+1\rangle$ at an \emph{effective virtual temperature}
\begin{equation}\label{VitualTemDef}
T_v=\frac{\hbar\omega_h-\hbar\omega_w}{\hbar\omega_h/T_h-\hbar\omega_w/T_w}.
\end{equation}
The cold mode at initial temperature $T_c$ then interacts with the virtual mode at $T_v$. In the regime $0\leq T_v< T_c$, cooling occurs since net energy is transferred from the cold mode to the virtual mode at a lower virtual temperature. At $T_v = T_c$, which corresponds to the equilibrium condition \eqref{stationaryCond}, no energy exchange takes place between the two modes.

\begin{figure}[t!]
   \centering
        \includegraphics[width=\columnwidth]{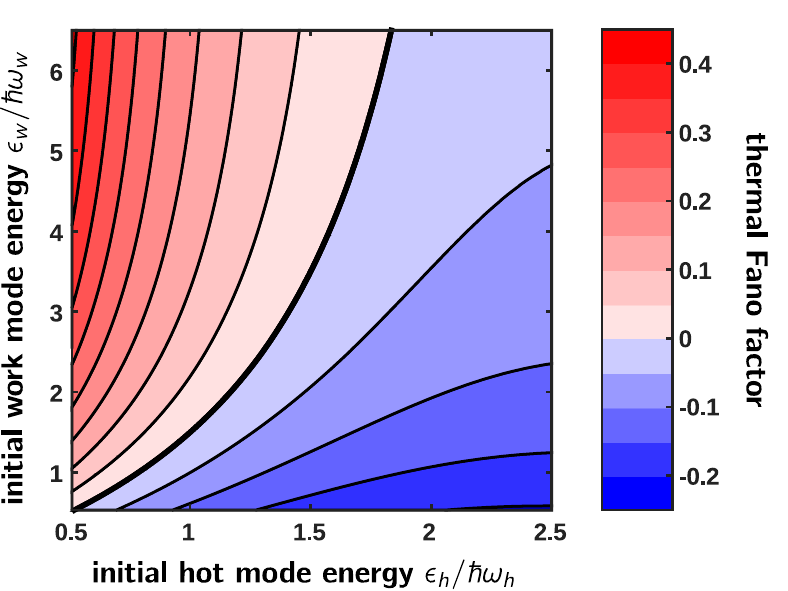}
        \caption{Thermal Fano factor \eqref{eq:Fano} of the infinite time-averaged state reduced to the cold mode for varied initial conditions as in Fig.~\ref{fig:changeInCmode}. A value of zero indicates a thermal energy distribution, i.e.~that the energy variance is that of a Gibbs state. We find that the variance is consistently above (below) thermal in the cooling (heating) regime.}
    \label{fig:varainceAthermal}
\end{figure}

Two important remarks are now in order: Firstly and unsurprisingly, the asymptotic three-mode state \eqref{infTimeAve}, which determines the long-time average energies after interaction, is in general not a tensor product of thermal states $\sigma\neq\rho_h^{\mathrm{th}} \otimes \rho_w^{\mathrm{th}} \otimes \rho_c^{\mathrm{th}}$. Secondly, the reduced single-mode states $\mathrm{tr}_{j\neq i}\{\sigma\}=\rho_i^{\mathrm{th}}$ do not exhibit a thermal energy distribution either. This can be easily illustrated by looking at the second moments of the local distributions. The deviation of, say, the cold mode energy variance in the time-averaged state $\sigma$ from that of a Gibbs distribution can be captured in the Fano factor 
\begin{equation}\label{eq:Fano}
 q = \frac{\braket{(\oa_c^\dagger \oa_c)^2} - \left( \braket{\oa_c^\dagger \oa_c} \right)^2}{\braket{\oa_c^\dagger \oa_c} \left( \braket{\oa_c^\dagger \oa_c} + 1 \right)} - 1.
\end{equation}
Fig.~\ref{fig:varainceAthermal} shows a contour plot of this quantity as a function of initial average energies as in Fig.~\ref{fig:changeInCmode}.
One sees that the variance is consistently higher ($q>0$) than that of a thermal state of the same energy in the cooling regime, and lower in the heating regime. At the same time, we found that the single-mode entropy is always lower than that of the thermal state, suggesting that the time-averaged energy distribution is more biased. 

Hence, and even though the reduced single-mode states remain diagonal at all times, we cannot assign temperatures to the asymptotic state $\sigma$ that the trilinear system approaches in the long-time average. This is in contrast to the three-qubit scenario, where one can formally associate a temperature to any diagonal single-qubit state \cite{Brask+Brunner:15, Mitchison+3:15}. Here, the three modes are correlated and explicitly driven out of thermal equilibrium by the trilinear interaction (which remains the case in the presence of simultaneous weak bath couplings).  This highlights the difference between genuine thermalization in open systems undergoing thermodynamic cycles and effective equilibration at an athermal state in quantum systems with non-trivial Hamiltonians \cite{Kulkarni+2:12, Gogolin+Eisert:16, Farrelly+2:17}.

\section{Enhanced cooling in the single-shot regime}\label{SingleShot}

One important signature of the cooling dynamics that can be seen from Fig.~\ref{fig:heating2cooling} is that the cold mode always overshoots to a lower-energy state at transient time scales, before it approaches the long-time average value. Hence, if one can control the interaction and halt the dynamics at the appropriate time, the cooling performance can be enhanced. 
A similar, measurement-induced transient cooling of a single qubit was found in Ref.~\cite{Erez2008}.
In the three-qubit scenario, this feature termed single-shot cooling was linked to the presence of quantum coherence \cite{Brask+Brunner:15, Mitchison+3:15}. Here, a similar effect can be observed as well \cite{Maslennikov+8:17}. The difference is that, in the three-qubit case, the system passes through many transient oscillations before reaching an equilibrium with help of the simultaneous thermalization with three independent baths. Here, the closed system alone approaches an effective equilibrium rather rapidly after the first overshooting oscillation. Irregular oscillations around the asymptotic state $\sigma$ then prevail, but at comparably small amplitudes. The higher the initial temperatures, the more negligible these oscillations become. 

Let us now examine the role of coherence in this transient effect. Following the qubit analogy \cite{Mitchison+3:15}, we compare the coherent energy exchange driven by the trilinear Hamiltonian to  incoherent implementations of the same exchange process. The simplest incoherent model is to assume the excitation of the hot mode $\oL = \oa_h^\dagger \oa_w \oa_c$ and the de-excitation $\oL^\dagger$ occur spontaneously at the same rate $\gamma$. The corresponding master equation in the interaction frame reads as
\begin{equation}\label{incoherent1}
\dot{\rho} = \gamma \left( \oL \rho \oL^\dagger + \oL^\dagger \rho \oL -\frac{1}{2} \left\{ \oL^\dagger \oL + \oL\oL^\dagger, \rho \right\} \right),
\end{equation}
which replaces the von Neumann equation with the trilinear Hamiltonian \eqref{Hinteraction} of the coherent model. One can easily check that the master equation still conserves both $N$ and $M$, and that it does not build up coherence between the combined Fock basis vectors $|n,N-n,M-n \rangle$; initially diagonal states remain diagonal at all times. The latter facilitates an efficient numerical implementation.

Alternatively, one can conceive a random unitary model where the coherent energy exchange driven by $\oH_{\rm int} \propto \oL + \oL^\dagger$ is switched on and off at random times with an average exchange rate $\gamma$,
\begin{equation}\label{incoherent2}
\dot{\rho} = \frac{\gamma}{2} \left[ \oL + \oL^\dagger, \left[ \rho, \oL + \oL^\dagger \right] \right].
\end{equation}
The master equation describes dephasing in the eigenbasis of the interaction Hamiltonian. The final equilibrium state is given by the fully dephased, or infinite time-averaged, initial state \eqref{infTimeAve}. Note however that the final state $\sigma$ does contain nondiagonal coherence terms when represented in terms of the $|n,N-n,M-n \rangle$ basis. At the same time, neither the coherent nor the two incoherent models generate coherence locally, i.e.~nondiagonal elements in the Fock state representation of the reduced single-mode states.

\begin{figure}[t!]
   \centering
        \includegraphics[width=\columnwidth]{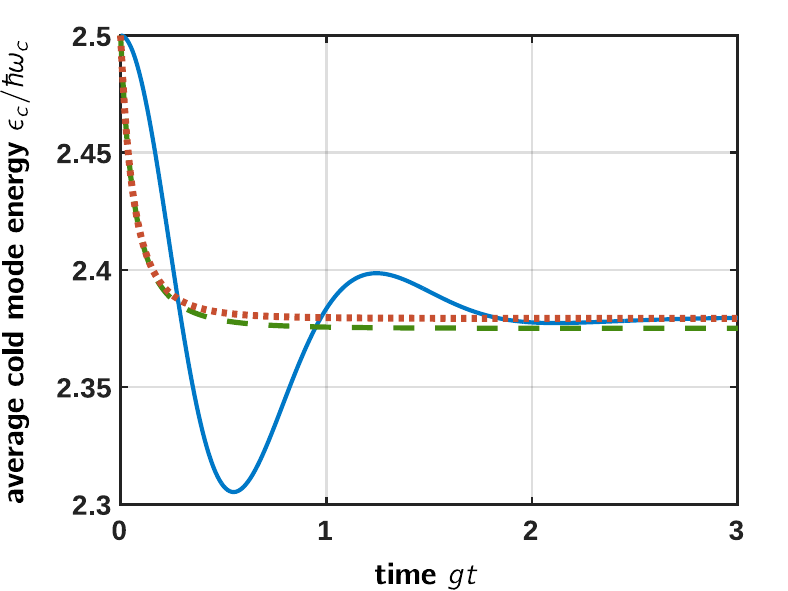}
        \caption{Comparison of coherent and incoherent time evolution for the average cold mode energy, starting from initial thermal energies $\eps_h(0)=1.5 \hbar\omega_h$, $\eps_w(0)=5.5 \hbar\omega_w$, and $\eps_c(0)=2.5 \hbar\omega_c$ (bottom panel in Fig.~\ref{fig:heating2cooling}). The solid line represents the coherent evolution, which exhibits a characteristic transient overshoot to below the long-time average for $gt < 1$. This behavior is not predicted by the incoherent models \eqref{incoherent1} and \eqref{incoherent2} for the same energy exchange between the modes, based on spontaneous excitation processes (dashed) or on phase-incoherent exchange (dotted), respectively. The latter yields the correct long-time average. We assume exchange rates equal to the coherent coupling constant $g$.}
    \label{fig:VSIncoherent}
\end{figure}

An exemplary comparison of the models in terms of the time evolution of the average cold mode energy is shown in Fig.~\ref{fig:VSIncoherent}, assuming $\gamma=g$ for simplicity. As expected, neither the fully incoherent model (green dashed line) nor the dephasing model (orange dotted) reproduce the enhanced cooling feature of the coherent model (blue solid). Instead, both incoherent models lead to an exponential decay towards their equilibrium values; the one of the dephasing model \eqref{incoherent2} is the value that the coherent evolution oscillates around and approaches in the long time average.

The time evolution and the equilibrium states differ slightly here, whereas in the previously studied three-qubit scenario there is no difference between the two incoherent models. This can be understood if we restrict to initial states $\rho(0)$ from the two-dimensional subspace $N=M=1$. The trilinear interaction is then equivalent to that of three qubits, and the two master equations \eqref{incoherent1} and \eqref{incoherent2} are identical.

Finally, we note that the comparison between the coherent and incoherent exchange process might suggest a quantum advantage in the transient cooling performance over ``classical'' implementations. This argument is based on the notion of ``classical'' states as a subset of quantum states without coherence in the relevant basis representation. In Section~\ref{classicalModel}, we provide an alternative view by comparing the coherent quantum model to its counterpart in the fully classical framework. It will turn out that the cooling enhancement is not a genuine quantum effect.
Indeed, the fact that the reduced single-mode quantum states always remain diagonal in the single-mode Fock basis suggests that the trilinear Hamiltonian only establishes coherence \emph{between} the modes. Such a notion of inter-mode coherence, in the sense of a finite interference capability and well-defined phase relation over a given coherence time, can also be understood in classical optics \cite{Glauber2006}.

\section{Effective Equilibration}\label{QuantumEquilibration}

A striking feature of the studied three-mode interaction, which we have observed and exploited in the previous sections, is the fast \textit{effective equilibration} of the average single-mode energies: even though the time evolution is strictly reversible, they appear to converge towards the stationary values of the time-averaged, fully dephased state, with only minor oscillations around those values for times $t \gg 1/g$. The phenomenon of effective equilibration has been explored in finite systems of high dimension \cite{Jensen+Shankar:85, Tasaki1998, Reimann:08, Rigol+2:08, Short:11,Ponomarev+2:11, Short+Farrelly:12, Reimann:12, Malabarba+4:14, Malabarba+2:15, Goldstein+2:15, Kaufman+6:16, Gogolin+Eisert:16, Pintos+4:15}, and is observed here for the case of initially thermal states. The effect is most pronounced at high initial temperatures, when the thermal occupation of the modes extends to high-dimensional subspaces associated to large values of the conserved quantities $N$ and $M$. A proper equilibration in the presence of a bath would lead to the same values, as long as the three-body coupling $g$ remains the dominant rate (see Appendix~\ref{OpenSystem}). This underpins the thermodynamical assessment based on the time-averaged steady state of the three modes as a model absorption refrigerator system in Sect.~\ref{AbsFridge}. 

\begin{figure*}[ht!]
   \centering
        \includegraphics[width=\textwidth]{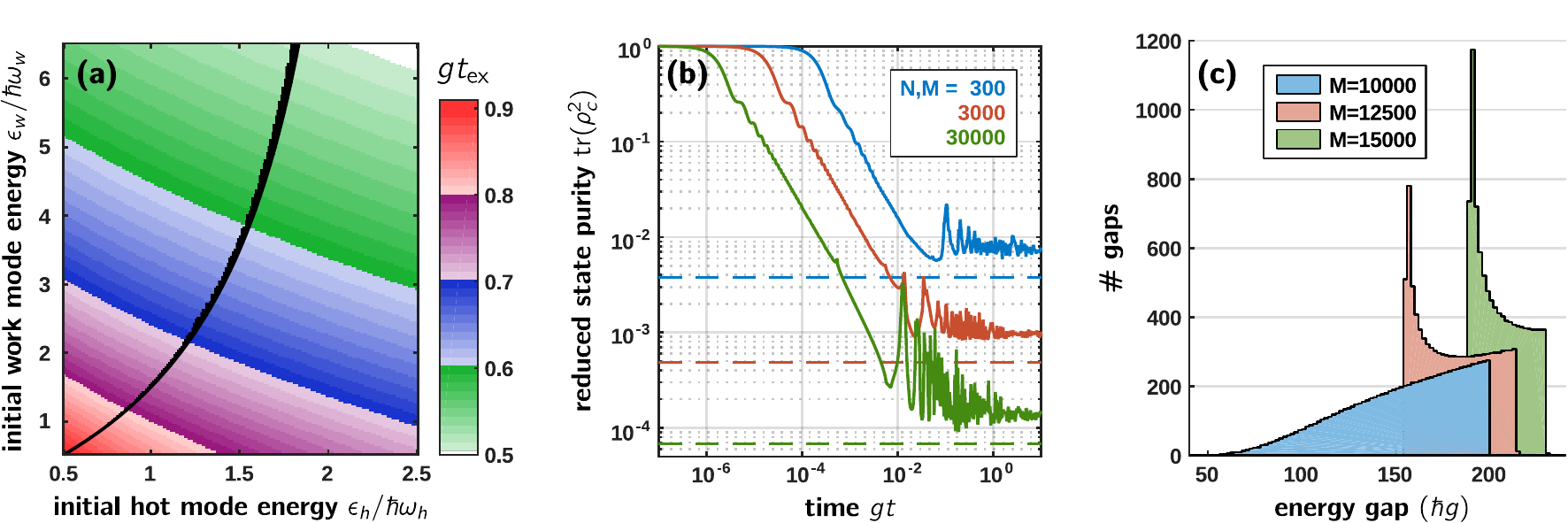}
\caption{\label{fig:quantumEquilibration}Features of effective equilibration:\quad (a) Time $gt_{\rm ex}$ of the transient overshoot extremum in the coherent evolution of the average mode energies for varied initial conditions as in Fig.~\ref{fig:changeInCmode}. The values were extracted from a numerical simulation in steps of $0.01/g$. The black pixels correspond to stationary states satisfying \eqref{stationaryCond}, where no such extremum is found. (b) Purity of the reduced single-mode state as a function of time when the unitary evolution starts from a pure triple Fock state $|100,N-100,M-100\rangle$ with varying values for $N=M$. Notice the double-logarithmic scale. The dashed lines represent the fully dephased states. Both the characteristic half-life time and the long-time averages of the purity scale in proportion to the inverse effective Hilbert subspace dimension $N+1$. (c) Exemplary energy gap spectra associated to the trilinear Hamiltonian in the subspaces of $N=10^4$ and varying $M$. The histograms are binned in units of $2\hbar g$.}
\end{figure*}

Quantitatively, one may speak of effective equilibration with respect to a set of observables if the cumulative time averages of their expectation values over the coherent time evolution quickly converge close to the values associated to a dephased stationary state \cite{Reimann:08,Linden2009,Short:11,Reimann2012a}. The characteristic convergence time scale and the residual difference should typically decrease with the (finite or effective) system dimension \cite{Short+Farrelly:12, Malabarba+4:14}. This implies that recurrences of the initial values, if present, must be short-lived and rare under unitary evolution. 

We observe such behaviour in the present, formally infinite-dimensional, scenario where the observables of interest are the single-mode energies. Our numerical studies show that both the characteristic equilibration time and the residual difference to the expectation values associated to the dephased state \eqref{infTimeAve} decrease with increasing initial occupation numbers $\bar{n}_i$. 
We could not observe recurrences in our simulations, some of which extended over time intervals a few orders of magnitude greater than $1/g$, see e.g.~Appendix~\ref{OpenSystem}.
In Fig.~\ref{fig:quantumEquilibration}(a), we plot the time at which the cold mode energy assumes its extremal overshoot value, which can serve as an estimate for the characteristic equilibration time. The times are extracted from unitary time evolutions with time increments of $0.01/g$ for varying initial conditions as in Fig.~\ref{fig:changeInCmode}. We observe that the values decrease with growing mode temperatures and effective Hilbert space dimension covered by the initial state, except at thermal equilibrium (black line).

We note that for pure initial Fock product states $|n,N-n,M-n \rangle$, one rather observes persistent strong and irregular oscillations of the average energies instead of the described equilibration. The latter is a consequence of averaging over the incommensurate sub-spectra of the trilinear Hamiltonian associated to its different orthogonal subspaces. Nevertheless, signatures of effective equilibration also exist at the level of quantum states in high-dimensional subspaces, e.g.~by looking at entangling dynamics \cite{Shaffer+3:14}.  As an example, we show in  Fig.~\ref{fig:quantumEquilibration}(b) how the reduced single-mode purity decays as a function of time for an initial Fock product state with $n=100$ at varying subspace dimension. The blue, red, and green lines (top to bottom) in the double-logarithmic plot correspond to $N=M=300$, $3000$, and $30000$, respectively. The dashed lines represent the reduced single-mode purities of the respective dephased states. 
As expected, the trilinear interaction rapidly entangles the three modes, and we observe a fast decay of single-mode purity close to the dephased values. Both these values and the characteristic half-life time decrease roughly in proportion to the inverse subspace dimension $d=\min \{N,M \} + 1$.

On the other hand, rapid equilibration in finite high-dimensional (e.g.~interacting many-body) systems has been linked to the non-Poissonian nature and spread of the underlying energy gap spectrum \cite{Sengupta2004,Manmana2007,Luitz2015,Gogolin+Eisert:16}, and it is assumed to appear in nonintegrable systems whose classical counterpart may exhibit signatures of chaos \cite{Kollath2007,Rigol2009,Banuls2011}. 

The energy gap spectrum of the quantum model varies strongly with the choice of the conserved quantities $(N,M)$, if restricted to a single invariant subspace. At large subspace dimension, non-Poissonian energy gap distributions are observed, in agreement with the observation of rapid equilibration. Three exemplary histograms of energy gaps (i.e.~differences between subsequent energy levels) are shown in Fig.~\ref{fig:quantumEquilibration}(c). They correspond to increasing values of $M$ at fixed $N=10000$ and subspace dimension $d=N+1$, using a bin width of $2\hbar g$. In all cases, the probability does not drop with the gap size in a Poisson-like fashion, as associated to non-equilibrating Hamiltonians.

Taken together, these intricate equilibration features distinguish the resonant three-mode Hamiltonian \eqref{HinteractionPicture} from other simple exchange coupling models. The resonant two-mode exchange coupling, for instance, can be diagonalized explicitly and leads merely to a normal mode splitting with a single fixed energy gap given by the coupling frequency. The same reversible dynamics can be seen in the two-dimensional subspaces (for $N=1$ or $M=1$) of the present system. At high initial excitations, the three-oscillator interaction model presented here could provide a minimal, accessible, and practically relevant testbed for studies of effective equilibration.

\section{Classical analysis}\label{classicalModel}

The present quantum refrigerator model of three trilinearly coupled harmonic oscillators admits direct benchmarking against the predictions of classical physics. We arrive at the classical version of the system straightforwardly by replacing the mode operators $\oa_i$ and $\oa^\dagger_i$ with canonical complex variables  $\alpha_i$ and $\alpha^{*}_i$. The Hamilton function becomes
\begin{equation}\label{classicalHint}
H = \sum_{i=h,w,c}\hbar \omega_i |\alpha_i|^2 +  \hbar g(\alpha_h\alpha^*_w\alpha^*_c+\alpha_h^*\alpha_w\alpha_c).
\end{equation}
The corresponding Hamilton equations of motion can be solved analytically after switching to the action-angle representation \cite{Bajer+Miranowicz:01}. For this, the complex amplitudes are expressed in terms of magnitude and phase, $\alpha_i=\sqrt{\iota_i}\mathrm{exp}(i\phi_i)$, where the $\iota_i$ represents the energy contained in mode $i$ in units of $\hbar \omega_i$.
As in the quantum case, one finds that the sums $I_1=\iota_h+\iota_w$ and $I_2=\iota_h+\iota_c$ are conserved, as well as the total phase difference $\Phi = \phi_w + \phi_c - \phi_h $.
Given the initial $\iota_i (0)$ and $\phi_i (0)$ and the additional constant of motion $L=\iota_h\iota_w\iota_c\mathrm{cos}^2(\Phi)$, this leads to the solution
\begin{align}\label{classicalSol}
\iota_h(t) &= c+(b-c)\mathrm{sn}^2 \left( \pm \sqrt{a-c}gt+\theta_0 \,|\, m \right), \\
 \theta_0 &= \text{sn}^{-1} \left( \sqrt{\frac{\iota_h (0)-c}{b-c}} \,\Big|\, m \right), \quad m = \frac{b-c}{a-c}. \nonumber
\end{align}
Here, $\mathrm{sn}( u \,|\, m )$ denotes the Jacobi elliptic function\footnote{We adopt the usual convention for the elliptic integrals \cite{Abramowitz1965}, which differs from the one in \cite{Bajer+Miranowicz:01}.} 
and $a>b>c$ are the three ordered solutions of the equations: $a+b+c=I_1+I_2$, $ab+bc+ca=I_1I_2$, and $abc=L$. 
The correct sign in the argument is given by that of $ \dot{\iota}_h (0) \propto - \sin \Phi(0)$.

The solution \eqref{classicalSol} is periodic with period $2K(m)$, where $K$ denotes a complete elliptic integral of the first kind. The time average of the energy contained in each mode is obtained by integrating \eqref{classicalSol} over one period, yielding a compact expression in terms of elliptic integrals of first and second kind
\begin{equation}\label{classicalSolTimeAvg}
\langle \iota_h \rangle_t = c + \frac{b-c}{m}\left [1-\frac{E(m)}{K(m)}\right].
\end{equation}
This is the classical counterpart of the quantum long time average, which is the mean energy associated to a completely dephased state.

For a comparison to the quantum simulation with thermal initial states, we evaluate the classical time evolution in a Monte-Carlo simulation by drawing initial angles $\phi_i (0)$ according to a uniform distribution and  $\iota_i (0)$ according to an exponential distribution with mean values $\bar{\iota}_i = k_B T_i/\hbar \omega_i$. 
However, the different energy statistics leaves room for a certain ambiguity in the matching of quantum and classical initial conditions. One can either match the initial temperatures $T_i$, arguing that the quantum and classical predictions shall be based on the same physical boundary conditions. Then the initial mean energies $\hbar \omega_i \bar{\iota}_i$ and $\eps_i = \hbar\omega_i (\bar{n}_i + 1/2)$ will differ slightly. Or one matches $\bar{\iota}_i = \bar{n}_i + 1/2$, which implies a better matching of the predicted energy trajectories at the cost of slightly different reservoir temperatures.

\begin{figure}[t!]
   \centering
        \includegraphics[width=\columnwidth]{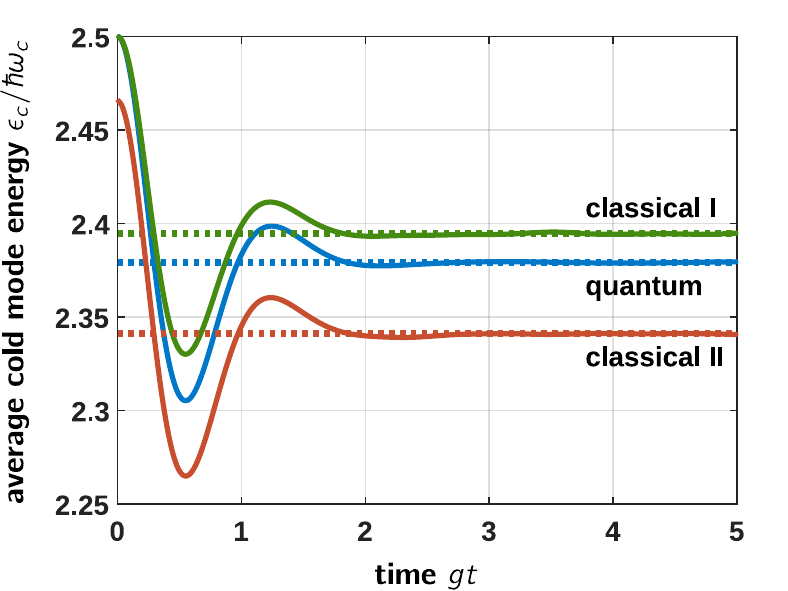}
        \caption{Comparison between the quantum and classical predictions for the time evolution of the average cold mode energy, starting from the initial conditions of Fig.~\ref{fig:VSIncoherent}. The classical results are obtained by a Monte-Carlo simulation of the solution \eqref{classicalSol} associated to the Hamilton function \eqref{classicalHint}, averaging over random samples of initial conditions. The two curves each correspond to $10^7$ trajectories drawn from a thermal distribution using either the same average mode energies (I, green) or the same initial temperatures (II, orange) as in the quantum case. The dotted lines indicate the infinite time averages.}
    \label{fig:classicalAnalysis}
\end{figure}

In Fig.~\ref{fig:classicalAnalysis}, we compare both classical options to the quantum prediction for an exemplary initial configuration in the cooling regime. The classical simulations were averaged over $10^7$ trajectories each. 
The relative difference to the quantum result decreases at higher initial temperatures where the continuous classical distribution of energies approximates well the discrete quantum statistics.

As expected, the classical result with matching initial energies (I) starts at the same point and remains closer to quantum case than the result with matching temperatures (II). 
Most importantly, the classical model predicts the same dynamical features as the quantum model, namely the short-time cooling enhancement and the effective equilibration at long times. 
The latter can again be attributed to thermal averaging: Even though single classical trajectories \eqref{classicalSol} describe periodic orbits, their periods depend on the initial energies and relative phases between the modes, and so they vary broadly over the thermal distribution of initial conditions. The averaged time evolution is not periodic and does not exhibit recurrences even for very long interaction times $gt \gg 1$.

In Section~\ref{SingleShot}, we have seen that the transient enhancement known as single-shot cooling is linked to the presence of coherence between eigenstates of the interaction Hamiltonian in the quantum framework. 
Now the quantum-classical comparison shows that this feature is not relying on genuinely quantum coherence, but is rather a generic feature of the particular three-mode exchange interaction. Indeed, it appears as well in the classical framework where coherence between harmonic modes exists in the form of a fixed phase relation between their oscillations.
The transient cooling enhancement can be suppressed by subjecting the classical phase coordinate $\Phi(t)$ to external noise, just as the feature will disappear in the quantum case if dephasing is present.

\section{Conclusion}\label{conclusion}

In this article, we have provided an in-depth study of a quantum absorption refrigerator model based on three resonantly coupled harmonic oscillator modes, as recently realized in an ion-trap experiment \cite{Maslennikov+8:17}. A closed-system analysis already exhibits the key features of the earlier studied analoguous three-qubit refrigerator model, in particular, the description of heating and cooling regimes in terms of virtual temperatures, and the enhanced cooling performance at transient time scales. 

On the other hand, we have found important differences and new insights in the more complex three-mode system. Even in the absence of simultaneous bath coupling, the initially thermal energy distributions of the cold, the hot, and the work modes equilibrate effectively around athermal distributions corresponding to a state that is fully dephased in the eigenbasis of the interaction Hamiltonian. This state exhibits residual coherences in the combined three-mode Fock representation, and it differs from the steady state of a fully incoherent implementation based on spontaneous excitation and de-excitation.

Most strikingly, we found that all essential features of the model can be explained by entirely classical means, with only minor quantitative differences that become irrelevant with growing temperatures. This includes, in particular, the enhanced transient cooling feature that had previously been attributed to quantum coherence in the three-qubit scenario. With the possibility of a one-to-one comparison between quantum and classical predictions at hand, the three-mode system may in fact serve as an ideal testbed to elucidate the role of quantum resources in thermodynamics.

\begin{acknowledgments}
We acknowledge clarifying discussions with Nicolas Brunner, Christian Gogolin, Alioscia Hamma, Sandu Popescu, Anthony Short, and Paul Skrzypczyk. 
This research is supported by the Ministry of Education, Singapore, and by the National Research Foundation, Prime Ministers Office, Singapore, under the Research Centres of Excellence Programme and through the Competitive Research Programme (Award No. NRF-CRP12-2013-03).
\end{acknowledgments}

\appendix
\numberwithin{equation}{section}

\section{Influence of simultaneous bath coupling}\label{OpenSystem}
Here we investigate the performance of the refrigerator when the three harmonic modes are simultaneously interacting via the trilinear Hamiltonian and each coupled to their individual thermal reservoirs. For that, we employ the master equation
\begin{equation} \label{eq:OpenFridge}
\frac{\text{d}\rho(t)}{\text{d}t}=-i[\oH_\text{int},\rho(t)]+\sum_{j=h,w,c}\mathcal{L}_j\rho(t),
\end{equation}
where $\mathcal{L}_j$ is the superoperator describing the effect of the bath coupled with mode $j$. We explicitly assume that each mode couples to its own bath locally and separately, i.e.~we employ the usual Lindblad dissipators, 
\begin{equation}\label{masterEqnLocal}
\mathcal{L}_j\rho=\kappa_j(1+\bar{n}_j)\mathcal{D}[\hat{a}_j]\rho+\kappa_j\bar{n}_j\mathcal{D}[\hat{a}_j^\dagger]\rho,
\end{equation}
with $\mathcal{D}[\oL]\rho=\oL\rho \oL^\dagger-\frac{1}{2}\{\oL^\dagger \oL,\rho\}$. Here $\kappa_j$ is the coupling strength of the mode $j$ with its bath and $\bar{n}_j$ is the mean occupation number at the bath temperature $T_j$. In the following, we assume that the bath couplings of the three modes are equal, $\kappa_h=\kappa_w=\kappa_c=\kappa$. 

It was pointed out that using local dissipators \eqref{masterEqnLocal} in a system of inter-coupled modes might not always yield consistent predictions for transient quantities such as the energy flow between the subsystems \cite{Levy+Kosloff:14,Gonzalez2017,Hofer2017}. However, since we are concerned here with the qualitative behavior of local observables (namely, single-mode energies), local dissipators may be used as an approximation.

In order to observe the refrigeration effect, the bath coupling $\kappa$ should not be too strong compared to the trilinear coupling $g$. Otherwise the modes will stay thermalized with their baths and cannot evolve away from the initial state. The system dynamics will be frozen. 

\begin{figure}[t!]
   \centering
        \includegraphics[width=\columnwidth]{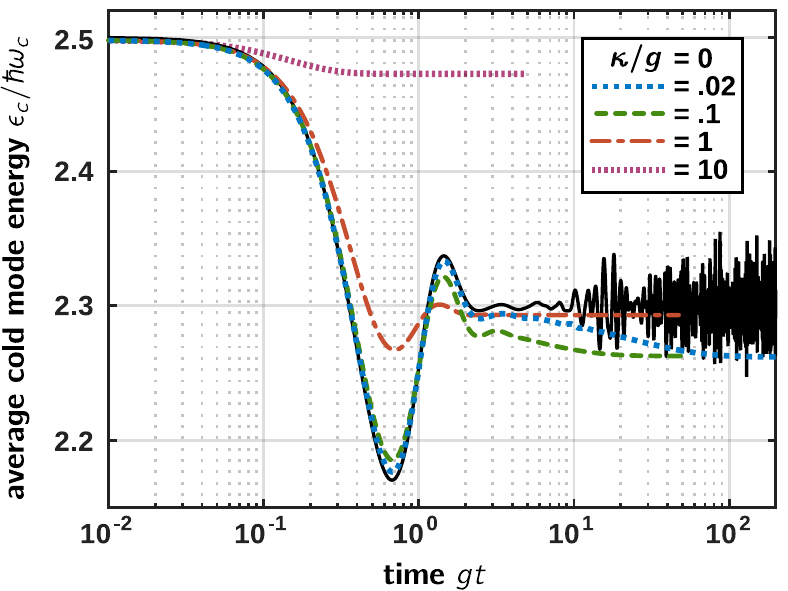}
        \caption{Time evolution of the average cold mode energy for different coupling rates $\kappa$ between the modes to their thermal baths. The open-system dynamics is described by the master equation \eqref{eq:OpenFridge}, which was simulated numerically using the QuTiP package, whereas $\kappa=0$ denotes the closed unitary evolution (solid line). We start from thermal initial states at $\eps_h (0) = \hbar\omega_h$, $\eps_w(0) = 3\hbar\omega_w$, and $\eps_c(0) = 2.5 \hbar\omega_c$.}
    \label{fig:OpenSystem}
\end{figure}

For $\kappa \ll g$ on the other hand, the dynamics is not expected to deviate much from the unitary evolution studied in the main text. Indeed, this is confirmed in Fig.~\ref{fig:OpenSystem}, which compares the evolution of the average cold mode energy for various thermalization rates to the unitary case. In particular, the transient oscillations at short times are well reproduced for $\kappa \ll g$ (green dashed and blue dotted line) and the steady-state value approached in the long-time limit does not differ significantly from the infinite time average of the unitary case (black solid line). 
We ran simulations up to $gt=10^4$ to confirm that deviations as significant as the initial oscillations do not recur.

Increasing the bath coupling $\kappa$ to the level of $g$ and beyond thwarts the coherent energy exchange between the modes; the transient oscillations disappear for $\kappa \geq g$ (red dash-dotted and purple fine-dotted line) and the mode energy quickly approaches equilibrium close to the initial value.

Concluding, the unitary dynamics studied in the main text approximates the transient refrigerator dynamics well in the relevant regime of operation, $\kappa \ll g$, and it gives reasonably good account of the long-time cooling performance.


\begin{thebibliography}{72}%
\makeatletter
\providecommand \@ifxundefined [1]{%
 \@ifx{#1\undefined}
}%
\providecommand \@ifnum [1]{%
 \ifnum #1\expandafter \@firstoftwo
 \else \expandafter \@secondoftwo
 \fi
}%
\providecommand \@ifx [1]{%
 \ifx #1\expandafter \@firstoftwo
 \else \expandafter \@secondoftwo
 \fi
}%
\providecommand \natexlab [1]{#1}%
\providecommand \enquote  [1]{``#1''}%
\providecommand \bibnamefont  [1]{#1}%
\providecommand \bibfnamefont [1]{#1}%
\providecommand \citenamefont [1]{#1}%
\providecommand \href@noop [0]{\@secondoftwo}%
\providecommand \href [0]{\begingroup \@sanitize@url \@href}%
\providecommand \@href[1]{\@@startlink{#1}\@@href}%
\providecommand \@@href[1]{\endgroup#1\@@endlink}%
\providecommand \@sanitize@url [0]{\catcode `\\12\catcode `\$12\catcode
  `\&12\catcode `\#12\catcode `\^12\catcode `\_12\catcode `\%12\relax}%
\providecommand \@@startlink[1]{}%
\providecommand \@@endlink[0]{}%
\providecommand \url  [0]{\begingroup\@sanitize@url \@url }%
\providecommand \@url [1]{\endgroup\@href {#1}{\urlprefix }}%
\providecommand \urlprefix  [0]{URL }%
\providecommand \Eprint [0]{\href }%
\providecommand \doibase [0]{http://dx.doi.org/}%
\providecommand \selectlanguage [0]{\@gobble}%
\providecommand \bibinfo  [0]{\@secondoftwo}%
\providecommand \bibfield  [0]{\@secondoftwo}%
\providecommand \translation [1]{[#1]}%
\providecommand \BibitemOpen [0]{}%
\providecommand \bibitemStop [0]{}%
\providecommand \bibitemNoStop [0]{.\EOS\space}%
\providecommand \EOS [0]{\spacefactor3000\relax}%
\providecommand \BibitemShut  [1]{\csname bibitem#1\endcsname}%
\let\auto@bib@innerbib\@empty
\bibitem [{\citenamefont {Horodecki}\ and\ \citenamefont
  {Oppenheim}(2013)}]{Horodecki+Oppenheim:13}%
  \BibitemOpen
  \bibfield  {author} {\bibinfo {author} {\bibfnamefont {M.}~\bibnamefont
  {Horodecki}}\ and\ \bibinfo {author} {\bibfnamefont {J.}~\bibnamefont
  {Oppenheim}},\ }\href {\doibase 10.1038/ncomms3059} {\bibfield  {journal}
  {\bibinfo  {journal} {Nat. Commun.}\ }\textbf {\bibinfo {volume} {4}},\
  \bibinfo {pages} {2059} (\bibinfo {year} {2013})}\BibitemShut {NoStop}%
\bibitem [{\citenamefont {Skrzypczyk}\ \emph {et~al.}(2014)\citenamefont
  {Skrzypczyk}, \citenamefont {Short},\ and\ \citenamefont
  {Popescu}}]{Skrzypczyk+2:14}%
  \BibitemOpen
  \bibfield  {author} {\bibinfo {author} {\bibfnamefont {P.}~\bibnamefont
  {Skrzypczyk}}, \bibinfo {author} {\bibfnamefont {A.~J.}\ \bibnamefont
  {Short}}, \ and\ \bibinfo {author} {\bibfnamefont {S.}~\bibnamefont
  {Popescu}},\ }\href {\doibase 10.1038/ncomms5185} {\bibfield  {journal}
  {\bibinfo  {journal} {Nat. Commun.}\ }\textbf {\bibinfo {volume} {5}},\
  \bibinfo {pages} {4185} (\bibinfo {year} {2014})}\BibitemShut {NoStop}%
\bibitem [{\citenamefont {Scully}\ \emph {et~al.}(2003)\citenamefont {Scully},
  \citenamefont {Zubairy}, \citenamefont {Agarwal},\ and\ \citenamefont
  {Walther}}]{Scully2003}%
  \BibitemOpen
  \bibfield  {author} {\bibinfo {author} {\bibfnamefont {M.~O.}\ \bibnamefont
  {Scully}}, \bibinfo {author} {\bibfnamefont {M.~S.}\ \bibnamefont {Zubairy}},
  \bibinfo {author} {\bibfnamefont {G.~S.}\ \bibnamefont {Agarwal}}, \ and\
  \bibinfo {author} {\bibfnamefont {H.}~\bibnamefont {Walther}},\ }\href
  {\doibase 10.1126/science.1078955} {\bibfield  {journal} {\bibinfo  {journal}
  {Science}\ }\textbf {\bibinfo {volume} {299}},\ \bibinfo {pages} {862}
  (\bibinfo {year} {2003})}\BibitemShut {NoStop}%
\bibitem [{\citenamefont {Dillenschneider}\ and\ \citenamefont
  {Lutz}(2009)}]{Dillenschneider2009}%
  \BibitemOpen
  \bibfield  {author} {\bibinfo {author} {\bibfnamefont {R.}~\bibnamefont
  {Dillenschneider}}\ and\ \bibinfo {author} {\bibfnamefont {E.}~\bibnamefont
  {Lutz}},\ }\href {\doibase 10.1209/0295-5075/88/50003} {\bibfield  {journal}
  {\bibinfo  {journal} {Europhys.~Lett.}\ }\textbf {\bibinfo {volume} {88}},\
  \bibinfo {pages} {50003} (\bibinfo {year} {2009})}\BibitemShut {NoStop}%
\bibitem [{\citenamefont {Ro{\ss}nagel}\ \emph {et~al.}(2014)\citenamefont
  {Ro{\ss}nagel}, \citenamefont {Abah}, \citenamefont {Schmidt-Kaler},
  \citenamefont {Singer},\ and\ \citenamefont {Lutz}}]{Rossnagel+4:14}%
  \BibitemOpen
  \bibfield  {author} {\bibinfo {author} {\bibfnamefont {J.}~\bibnamefont
  {Ro{\ss}nagel}}, \bibinfo {author} {\bibfnamefont {O.}~\bibnamefont {Abah}},
  \bibinfo {author} {\bibfnamefont {F.}~\bibnamefont {Schmidt-Kaler}}, \bibinfo
  {author} {\bibfnamefont {K.}~\bibnamefont {Singer}}, \ and\ \bibinfo {author}
  {\bibfnamefont {E.}~\bibnamefont {Lutz}},\ }\href {\doibase
  10.1103/PhysRevLett.112.030602} {\bibfield  {journal} {\bibinfo  {journal}
  {Phys. Rev. Lett.}\ }\textbf {\bibinfo {volume} {112}},\ \bibinfo {pages}
  {030602} (\bibinfo {year} {2014})}\BibitemShut {NoStop}%
\bibitem [{\citenamefont {Correa}\ \emph {et~al.}(2014)\citenamefont {Correa},
  \citenamefont {Palao}, \citenamefont {Alonso},\ and\ \citenamefont
  {Adesso}}]{Correa+3:14}%
  \BibitemOpen
  \bibfield  {author} {\bibinfo {author} {\bibfnamefont {L.~A.}\ \bibnamefont
  {Correa}}, \bibinfo {author} {\bibfnamefont {J.~P.}\ \bibnamefont {Palao}},
  \bibinfo {author} {\bibfnamefont {D.}~\bibnamefont {Alonso}}, \ and\ \bibinfo
  {author} {\bibfnamefont {G.}~\bibnamefont {Adesso}},\ }\href {\doibase
  10.1038/srep03949} {\bibfield  {journal} {\bibinfo  {journal} {Sci. Rep.}\
  }\textbf {\bibinfo {volume} {4}},\ \bibinfo {pages} {3949} (\bibinfo {year}
  {2014})}\BibitemShut {NoStop}%
\bibitem [{\citenamefont {Strasberg}\ \emph {et~al.}(2017)\citenamefont
  {Strasberg}, \citenamefont {Schaller}, \citenamefont {Brandes},\ and\
  \citenamefont {Esposito}}]{Strasberg2017}%
  \BibitemOpen
  \bibfield  {author} {\bibinfo {author} {\bibfnamefont {P.}~\bibnamefont
  {Strasberg}}, \bibinfo {author} {\bibfnamefont {G.}~\bibnamefont {Schaller}},
  \bibinfo {author} {\bibfnamefont {T.}~\bibnamefont {Brandes}}, \ and\
  \bibinfo {author} {\bibfnamefont {M.}~\bibnamefont {Esposito}},\ }\href
  {\doibase 10.1103/PhysRevX.7.021003} {\bibfield  {journal} {\bibinfo
  {journal} {Phys. Rev. X}\ }\textbf {\bibinfo {volume} {7}},\ \bibinfo {pages}
  {021003} (\bibinfo {year} {2017})}\BibitemShut {NoStop}%
\bibitem [{\citenamefont {Niedenzu}\ \emph {et~al.}(2017)\citenamefont
  {Niedenzu}, \citenamefont {Mukherjee}, \citenamefont {Ghosh}, \citenamefont
  {Kofman},\ and\ \citenamefont {Kurizki}}]{Niedenzu2017}%
  \BibitemOpen
  \bibfield  {author} {\bibinfo {author} {\bibfnamefont {W.}~\bibnamefont
  {Niedenzu}}, \bibinfo {author} {\bibfnamefont {V.}~\bibnamefont {Mukherjee}},
  \bibinfo {author} {\bibfnamefont {A.}~\bibnamefont {Ghosh}}, \bibinfo
  {author} {\bibfnamefont {A.~G.}\ \bibnamefont {Kofman}}, \ and\ \bibinfo
  {author} {\bibfnamefont {G.}~\bibnamefont {Kurizki}},\ }\href
  {http://arxiv.org/abs/1703.02911} {\  (\bibinfo {year} {2017})},\ \Eprint
  {http://arxiv.org/abs/1703.02911} {arXiv:1703.02911} \BibitemShut {NoStop}%
\bibitem [{\citenamefont {Allahverdyan}\ \emph {et~al.}(2004)\citenamefont
  {Allahverdyan}, \citenamefont {Balian},\ and\ \citenamefont
  {Nieuwenhuizen}}]{Allahverdyan2004}%
  \BibitemOpen
  \bibfield  {author} {\bibinfo {author} {\bibfnamefont {A.~E.}\ \bibnamefont
  {Allahverdyan}}, \bibinfo {author} {\bibfnamefont {R.}~\bibnamefont
  {Balian}}, \ and\ \bibinfo {author} {\bibfnamefont {T.~M.}\ \bibnamefont
  {Nieuwenhuizen}},\ }\href {\doibase 10.1209/epl/i2004-10101-2} {\bibfield
  {journal} {\bibinfo  {journal} {Europhys.~Lett.}\ }\textbf {\bibinfo {volume}
  {67}},\ \bibinfo {pages} {565} (\bibinfo {year} {2004})}\BibitemShut
  {NoStop}%
\bibitem [{\citenamefont {Talkner}\ \emph {et~al.}(2007)\citenamefont
  {Talkner}, \citenamefont {Lutz},\ and\ \citenamefont
  {H{\"{a}}nggi}}]{Talkner2007}%
  \BibitemOpen
  \bibfield  {author} {\bibinfo {author} {\bibfnamefont {P.}~\bibnamefont
  {Talkner}}, \bibinfo {author} {\bibfnamefont {E.}~\bibnamefont {Lutz}}, \
  and\ \bibinfo {author} {\bibfnamefont {P.}~\bibnamefont {H{\"{a}}nggi}},\
  }\href {\doibase 10.1103/PhysRevE.75.050102} {\bibfield  {journal} {\bibinfo
  {journal} {Phys. Rev. E}\ }\textbf {\bibinfo {volume} {75}},\ \bibinfo
  {pages} {050102} (\bibinfo {year} {2007})}\BibitemShut {NoStop}%
\bibitem [{\citenamefont {Campisi}\ \emph {et~al.}(2011)\citenamefont
  {Campisi}, \citenamefont {H{\"{a}}nggi},\ and\ \citenamefont
  {Talkner}}]{Campisi2011}%
  \BibitemOpen
  \bibfield  {author} {\bibinfo {author} {\bibfnamefont {M.}~\bibnamefont
  {Campisi}}, \bibinfo {author} {\bibfnamefont {P.}~\bibnamefont
  {H{\"{a}}nggi}}, \ and\ \bibinfo {author} {\bibfnamefont {P.}~\bibnamefont
  {Talkner}},\ }\href {\doibase 10.1103/RevModPhys.83.771} {\bibfield
  {journal} {\bibinfo  {journal} {Rev. Mod. Phys.}\ }\textbf {\bibinfo {volume}
  {83}},\ \bibinfo {pages} {771} (\bibinfo {year} {2011})}\BibitemShut
  {NoStop}%
\bibitem [{\citenamefont {Roncaglia}\ \emph {et~al.}(2014)\citenamefont
  {Roncaglia}, \citenamefont {Cerisola},\ and\ \citenamefont
  {Paz}}]{Roncaglia+2:14}%
  \BibitemOpen
  \bibfield  {author} {\bibinfo {author} {\bibfnamefont {A.~J.}\ \bibnamefont
  {Roncaglia}}, \bibinfo {author} {\bibfnamefont {F.}~\bibnamefont {Cerisola}},
  \ and\ \bibinfo {author} {\bibfnamefont {J.~P.}\ \bibnamefont {Paz}},\ }\href
  {\doibase 10.1103/PhysRevLett.113.250601} {\bibfield  {journal} {\bibinfo
  {journal} {Phys. Rev. Lett.}\ }\textbf {\bibinfo {volume} {113}},\ \bibinfo
  {pages} {250601} (\bibinfo {year} {2014})}\BibitemShut {NoStop}%
\bibitem [{\citenamefont {Talkner}\ and\ \citenamefont
  {H{\"{a}}nggi}(2016)}]{Talkner+Hanggi:16}%
  \BibitemOpen
  \bibfield  {author} {\bibinfo {author} {\bibfnamefont {P.}~\bibnamefont
  {Talkner}}\ and\ \bibinfo {author} {\bibfnamefont {P.}~\bibnamefont
  {H{\"{a}}nggi}},\ }\href {\doibase 10.1103/PhysRevE.93.022131} {\bibfield
  {journal} {\bibinfo  {journal} {Phys. Rev. E}\ }\textbf {\bibinfo {volume}
  {93}},\ \bibinfo {pages} {022131} (\bibinfo {year} {2016})}\BibitemShut
  {NoStop}%
\bibitem [{\citenamefont {Geva}\ and\ \citenamefont
  {Kosloff}(1996)}]{Geva1996}%
  \BibitemOpen
  \bibfield  {author} {\bibinfo {author} {\bibfnamefont {E.}~\bibnamefont
  {Geva}}\ and\ \bibinfo {author} {\bibfnamefont {R.}~\bibnamefont {Kosloff}},\
  }\href {\doibase 10.1063/1.471453} {\bibfield  {journal} {\bibinfo  {journal}
  {J. Chem. Phys.}\ }\textbf {\bibinfo {volume} {104}},\ \bibinfo {pages}
  {7681} (\bibinfo {year} {1996})}\BibitemShut {NoStop}%
\bibitem [{\citenamefont {Venturelli}\ \emph {et~al.}(2013)\citenamefont
  {Venturelli}, \citenamefont {Fazio},\ and\ \citenamefont
  {Giovannetti}}]{Venturelli+2:13}%
  \BibitemOpen
  \bibfield  {author} {\bibinfo {author} {\bibfnamefont {D.}~\bibnamefont
  {Venturelli}}, \bibinfo {author} {\bibfnamefont {R.}~\bibnamefont {Fazio}}, \
  and\ \bibinfo {author} {\bibfnamefont {V.}~\bibnamefont {Giovannetti}},\
  }\href {\doibase 10.1103/PhysRevLett.110.256801} {\bibfield  {journal}
  {\bibinfo  {journal} {Phys. Rev. Lett.}\ }\textbf {\bibinfo {volume} {110}},\
  \bibinfo {pages} {256801} (\bibinfo {year} {2013})}\BibitemShut {NoStop}%
\bibitem [{\citenamefont {Kosloff}\ and\ \citenamefont
  {Levy}(2014)}]{Kosloff+Levy:14}%
  \BibitemOpen
  \bibfield  {author} {\bibinfo {author} {\bibfnamefont {R.}~\bibnamefont
  {Kosloff}}\ and\ \bibinfo {author} {\bibfnamefont {A.}~\bibnamefont {Levy}},\
  }\href {\doibase 10.1146/annurev-physchem-040513-103724} {\bibfield
  {journal} {\bibinfo  {journal} {Annu. Rev. Phys. Chem.}\ }\textbf {\bibinfo
  {volume} {65}},\ \bibinfo {pages} {365} (\bibinfo {year} {2014})}\BibitemShut
  {NoStop}%
\bibitem [{\citenamefont {Mari}\ \emph {et~al.}(2015)\citenamefont {Mari},
  \citenamefont {Farace},\ and\ \citenamefont {Giovannetti}}]{Mari+2:15}%
  \BibitemOpen
  \bibfield  {author} {\bibinfo {author} {\bibfnamefont {A.}~\bibnamefont
  {Mari}}, \bibinfo {author} {\bibfnamefont {A.}~\bibnamefont {Farace}}, \ and\
  \bibinfo {author} {\bibfnamefont {V.}~\bibnamefont {Giovannetti}},\ }\href
  {\doibase 10.1088/0953-4075/48/17/175501} {\bibfield  {journal} {\bibinfo
  {journal} {J. Phys. B}\ }\textbf {\bibinfo {volume} {48}},\ \bibinfo {pages}
  {175501} (\bibinfo {year} {2015})}\BibitemShut {NoStop}%
\bibitem [{\citenamefont {Hofer}\ \emph
  {et~al.}(2016{\natexlab{a}})\citenamefont {Hofer}, \citenamefont {Souquet},\
  and\ \citenamefont {Clerk}}]{Hofer+2:16}%
  \BibitemOpen
  \bibfield  {author} {\bibinfo {author} {\bibfnamefont {P.~P.}\ \bibnamefont
  {Hofer}}, \bibinfo {author} {\bibfnamefont {J.-R.}\ \bibnamefont {Souquet}},
  \ and\ \bibinfo {author} {\bibfnamefont {A.~A.}\ \bibnamefont {Clerk}},\
  }\href {\doibase 10.1103/PhysRevB.93.041418} {\bibfield  {journal} {\bibinfo
  {journal} {Phys. Rev. B}\ }\textbf {\bibinfo {volume} {93}},\ \bibinfo
  {pages} {041418} (\bibinfo {year} {2016}{\natexlab{a}})}\BibitemShut
  {NoStop}%
\bibitem [{\citenamefont {Joulain}\ \emph {et~al.}(2016)\citenamefont
  {Joulain}, \citenamefont {Drevillon}, \citenamefont {Ezzahri},\ and\
  \citenamefont {Ordonez-Miranda}}]{Joulain+3:16}%
  \BibitemOpen
  \bibfield  {author} {\bibinfo {author} {\bibfnamefont {K.}~\bibnamefont
  {Joulain}}, \bibinfo {author} {\bibfnamefont {J.}~\bibnamefont {Drevillon}},
  \bibinfo {author} {\bibfnamefont {Y.}~\bibnamefont {Ezzahri}}, \ and\
  \bibinfo {author} {\bibfnamefont {J.}~\bibnamefont {Ordonez-Miranda}},\
  }\href {\doibase 10.1103/PhysRevLett.116.200601} {\bibfield  {journal}
  {\bibinfo  {journal} {Phys. Rev. Lett.}\ }\textbf {\bibinfo {volume} {116}},\
  \bibinfo {pages} {200601} (\bibinfo {year} {2016})}\BibitemShut {NoStop}%
\bibitem [{\citenamefont {Mitchison}\ \emph {et~al.}(2016)\citenamefont
  {Mitchison}, \citenamefont {Huber}, \citenamefont {Prior}, \citenamefont
  {Woods},\ and\ \citenamefont {Plenio}}]{Mitchison+4:16}%
  \BibitemOpen
  \bibfield  {author} {\bibinfo {author} {\bibfnamefont {M.~T.}\ \bibnamefont
  {Mitchison}}, \bibinfo {author} {\bibfnamefont {M.}~\bibnamefont {Huber}},
  \bibinfo {author} {\bibfnamefont {J.}~\bibnamefont {Prior}}, \bibinfo
  {author} {\bibfnamefont {M.~P.}\ \bibnamefont {Woods}}, \ and\ \bibinfo
  {author} {\bibfnamefont {M.~B.}\ \bibnamefont {Plenio}},\ }\href {\doibase
  10.1088/2058-9565/1/1/015001} {\bibfield  {journal} {\bibinfo  {journal}
  {Quantum Sci. Technol.}\ }\textbf {\bibinfo {volume} {1}},\ \bibinfo {pages}
  {015001} (\bibinfo {year} {2016})}\BibitemShut {NoStop}%
\bibitem [{\citenamefont {Hofer}\ \emph
  {et~al.}(2016{\natexlab{b}})\citenamefont {Hofer}, \citenamefont
  {Perarnau-Llobet}, \citenamefont {Brask}, \citenamefont {Silva},
  \citenamefont {Huber},\ and\ \citenamefont {Brunner}}]{Hofer+5:16}%
  \BibitemOpen
  \bibfield  {author} {\bibinfo {author} {\bibfnamefont {P.~P.}\ \bibnamefont
  {Hofer}}, \bibinfo {author} {\bibfnamefont {M.}~\bibnamefont
  {Perarnau-Llobet}}, \bibinfo {author} {\bibfnamefont {J.~B.}\ \bibnamefont
  {Brask}}, \bibinfo {author} {\bibfnamefont {R.}~\bibnamefont {Silva}},
  \bibinfo {author} {\bibfnamefont {M.}~\bibnamefont {Huber}}, \ and\ \bibinfo
  {author} {\bibfnamefont {N.}~\bibnamefont {Brunner}},\ }\href {\doibase
  10.1103/PhysRevB.94.235420} {\bibfield  {journal} {\bibinfo  {journal} {Phys.
  Rev. B}\ }\textbf {\bibinfo {volume} {94}},\ \bibinfo {pages} {235420}
  (\bibinfo {year} {2016}{\natexlab{b}})}\BibitemShut {NoStop}%
\bibitem [{\citenamefont {Karimi}\ and\ \citenamefont
  {Pekola}(2016)}]{Karimi+Pekola:16}%
  \BibitemOpen
  \bibfield  {author} {\bibinfo {author} {\bibfnamefont {B.}~\bibnamefont
  {Karimi}}\ and\ \bibinfo {author} {\bibfnamefont {J.~P.}\ \bibnamefont
  {Pekola}},\ }\href {\doibase 10.1103/PhysRevB.94.184503} {\bibfield
  {journal} {\bibinfo  {journal} {Phys. Rev. B}\ }\textbf {\bibinfo {volume}
  {94}},\ \bibinfo {pages} {184503} (\bibinfo {year} {2016})}\BibitemShut
  {NoStop}%
\bibitem [{\citenamefont {Roulet}\ \emph {et~al.}(2017)\citenamefont {Roulet},
  \citenamefont {Nimmrichter}, \citenamefont {Arrazola}, \citenamefont {Seah},\
  and\ \citenamefont {Scarani}}]{Roulet+3:16}%
  \BibitemOpen
  \bibfield  {author} {\bibinfo {author} {\bibfnamefont {A.}~\bibnamefont
  {Roulet}}, \bibinfo {author} {\bibfnamefont {S.}~\bibnamefont {Nimmrichter}},
  \bibinfo {author} {\bibfnamefont {J.}~\bibnamefont {Arrazola}}, \bibinfo
  {author} {\bibfnamefont {S.}~\bibnamefont {Seah}}, \ and\ \bibinfo {author}
  {\bibfnamefont {V.}~\bibnamefont {Scarani}},\ }\href {\doibase
  10.1103/PhysRevE.95.062131} {\bibfield  {journal} {\bibinfo  {journal} {Phys.
  Rev. E}\ }\textbf {\bibinfo {volume} {95}},\ \bibinfo {pages} {062131}
  (\bibinfo {year} {2017})}\BibitemShut {NoStop}%
\bibitem [{\citenamefont {Bissbort}\ \emph {et~al.}(2017)\citenamefont
  {Bissbort}, \citenamefont {Teo}, \citenamefont {Guo}, \citenamefont {Casati},
  \citenamefont {Benenti},\ and\ \citenamefont {Poletti}}]{Bissbort+5:16}%
  \BibitemOpen
  \bibfield  {author} {\bibinfo {author} {\bibfnamefont {U.}~\bibnamefont
  {Bissbort}}, \bibinfo {author} {\bibfnamefont {C.}~\bibnamefont {Teo}},
  \bibinfo {author} {\bibfnamefont {C.}~\bibnamefont {Guo}}, \bibinfo {author}
  {\bibfnamefont {G.}~\bibnamefont {Casati}}, \bibinfo {author} {\bibfnamefont
  {G.}~\bibnamefont {Benenti}}, \ and\ \bibinfo {author} {\bibfnamefont
  {D.}~\bibnamefont {Poletti}},\ }\href {\doibase 10.1103/PhysRevE.95.062143}
  {\bibfield  {journal} {\bibinfo  {journal} {Phys. Rev. E}\ }\textbf {\bibinfo
  {volume} {95}},\ \bibinfo {pages} {062143} (\bibinfo {year}
  {2017})}\BibitemShut {NoStop}%
\bibitem [{\citenamefont {Zhang}\ and\ \citenamefont
  {Zhang}(2017)}]{Zhang2017}%
  \BibitemOpen
  \bibfield  {author} {\bibinfo {author} {\bibfnamefont {K.}~\bibnamefont
  {Zhang}}\ and\ \bibinfo {author} {\bibfnamefont {W.}~\bibnamefont {Zhang}},\
  }\href {\doibase 10.1103/PhysRevA.95.053870} {\bibfield  {journal} {\bibinfo
  {journal} {Phys. Rev. A}\ }\textbf {\bibinfo {volume} {95}},\ \bibinfo
  {pages} {053870} (\bibinfo {year} {2017})}\BibitemShut {NoStop}%
\bibitem [{\citenamefont {Reid}\ \emph {et~al.}(2017)\citenamefont {Reid},
  \citenamefont {Pigeon}, \citenamefont {Antezza},\ and\ \citenamefont {{De
  Chiara}}}]{Reid2017}%
  \BibitemOpen
  \bibfield  {author} {\bibinfo {author} {\bibfnamefont {B.}~\bibnamefont
  {Reid}}, \bibinfo {author} {\bibfnamefont {S.}~\bibnamefont {Pigeon}},
  \bibinfo {author} {\bibfnamefont {M.}~\bibnamefont {Antezza}}, \ and\
  \bibinfo {author} {\bibfnamefont {G.}~\bibnamefont {{De Chiara}}},\ }\href
  {http://arxiv.org/abs/1708.07435} {\  (\bibinfo {year} {2017})},\ \Eprint
  {http://arxiv.org/abs/1708.07435} {arXiv:1708.07435} \BibitemShut {NoStop}%
\bibitem [{\citenamefont {Hardal}\ \emph {et~al.}(2017)\citenamefont {Hardal},
  \citenamefont {Aslan}, \citenamefont {Wilson},\ and\ \citenamefont
  {M{\"{u}}stecaplıoğlu}}]{Hardal2017}%
  \BibitemOpen
  \bibfield  {author} {\bibinfo {author} {\bibfnamefont {A.~{\"{U}}.~C.}\
  \bibnamefont {Hardal}}, \bibinfo {author} {\bibfnamefont {N.}~\bibnamefont
  {Aslan}}, \bibinfo {author} {\bibfnamefont {C.~M.}\ \bibnamefont {Wilson}}, \
  and\ \bibinfo {author} {\bibfnamefont {{\"{O}}.~E.}\ \bibnamefont
  {M{\"{u}}stecaplıoğlu}},\ }\href {http://arxiv.org/abs/1708.01182} {\
  (\bibinfo {year} {2017})},\ \Eprint {http://arxiv.org/abs/1708.01182}
  {arXiv:1708.01182} \BibitemShut {NoStop}%
\bibitem [{\citenamefont {Mu}\ \emph {et~al.}(2017)\citenamefont {Mu},
  \citenamefont {Agarwalla}, \citenamefont {Schaller},\ and\ \citenamefont
  {Segal}}]{Mu2017}%
  \BibitemOpen
  \bibfield  {author} {\bibinfo {author} {\bibfnamefont {A.}~\bibnamefont
  {Mu}}, \bibinfo {author} {\bibfnamefont {B.~K.}\ \bibnamefont {Agarwalla}},
  \bibinfo {author} {\bibfnamefont {G.}~\bibnamefont {Schaller}}, \ and\
  \bibinfo {author} {\bibfnamefont {D.}~\bibnamefont {Segal}},\ }\href
  {http://arxiv.org/abs/1709.02835} {\  (\bibinfo {year} {2017})},\ \Eprint
  {http://arxiv.org/abs/1709.02835} {arXiv:1709.02835} \BibitemShut {NoStop}%
\bibitem [{\citenamefont {Palao}\ \emph {et~al.}(2001)\citenamefont {Palao},
  \citenamefont {Kosloff},\ and\ \citenamefont {Gordon}}]{Palao2001}%
  \BibitemOpen
  \bibfield  {author} {\bibinfo {author} {\bibfnamefont {J.~P.}\ \bibnamefont
  {Palao}}, \bibinfo {author} {\bibfnamefont {R.}~\bibnamefont {Kosloff}}, \
  and\ \bibinfo {author} {\bibfnamefont {J.~M.}\ \bibnamefont {Gordon}},\
  }\href {\doibase 10.1103/PhysRevE.64.056130} {\bibfield  {journal} {\bibinfo
  {journal} {Phys. Rev. E}\ }\textbf {\bibinfo {volume} {64}},\ \bibinfo
  {pages} {056130} (\bibinfo {year} {2001})}\BibitemShut {NoStop}%
\bibitem [{\citenamefont {Linden}\ \emph {et~al.}(2010)\citenamefont {Linden},
  \citenamefont {Popescu},\ and\ \citenamefont {Skrzypczyk}}]{Linden+2:10}%
  \BibitemOpen
  \bibfield  {author} {\bibinfo {author} {\bibfnamefont {N.}~\bibnamefont
  {Linden}}, \bibinfo {author} {\bibfnamefont {S.}~\bibnamefont {Popescu}}, \
  and\ \bibinfo {author} {\bibfnamefont {P.}~\bibnamefont {Skrzypczyk}},\
  }\href {\doibase 10.1103/PhysRevLett.105.130401} {\bibfield  {journal}
  {\bibinfo  {journal} {Phys. Rev. Lett.}\ }\textbf {\bibinfo {volume} {105}},\
  \bibinfo {pages} {130401} (\bibinfo {year} {2010})}\BibitemShut {NoStop}%
\bibitem [{\citenamefont {Levy}\ and\ \citenamefont
  {Kosloff}(2012)}]{Levy+Kosloff:12}%
  \BibitemOpen
  \bibfield  {author} {\bibinfo {author} {\bibfnamefont {A.}~\bibnamefont
  {Levy}}\ and\ \bibinfo {author} {\bibfnamefont {R.}~\bibnamefont {Kosloff}},\
  }\href {\doibase 10.1103/PhysRevLett.108.070604} {\bibfield  {journal}
  {\bibinfo  {journal} {Phys. Rev. Lett.}\ }\textbf {\bibinfo {volume} {108}},\
  \bibinfo {pages} {070604} (\bibinfo {year} {2012})}\BibitemShut {NoStop}%
\bibitem [{\citenamefont {Goold}\ \emph {et~al.}(2016)\citenamefont {Goold},
  \citenamefont {Huber}, \citenamefont {Riera}, \citenamefont {del Rio},\ and\
  \citenamefont {Skrzypczyk}}]{Goold+4:16}%
  \BibitemOpen
  \bibfield  {author} {\bibinfo {author} {\bibfnamefont {J.}~\bibnamefont
  {Goold}}, \bibinfo {author} {\bibfnamefont {M.}~\bibnamefont {Huber}},
  \bibinfo {author} {\bibfnamefont {A.}~\bibnamefont {Riera}}, \bibinfo
  {author} {\bibfnamefont {L.}~\bibnamefont {del Rio}}, \ and\ \bibinfo
  {author} {\bibfnamefont {P.}~\bibnamefont {Skrzypczyk}},\ }\href {\doibase
  10.1088/1751-8113/49/14/143001} {\bibfield  {journal} {\bibinfo  {journal}
  {J. Phys. A Math. Theor.}\ }\textbf {\bibinfo {volume} {49}},\ \bibinfo
  {pages} {143001} (\bibinfo {year} {2016})}\BibitemShut {NoStop}%
\bibitem [{\citenamefont {Maslennikov}\ \emph {et~al.}(2017)\citenamefont
  {Maslennikov}, \citenamefont {Ding}, \citenamefont {Hablutzel}, \citenamefont
  {Gan}, \citenamefont {Roulet}, \citenamefont {Nimmrichter}, \citenamefont
  {Dai}, \citenamefont {Scarani},\ and\ \citenamefont
  {Matsukevich}}]{Maslennikov+8:17}%
  \BibitemOpen
  \bibfield  {author} {\bibinfo {author} {\bibfnamefont {G.}~\bibnamefont
  {Maslennikov}}, \bibinfo {author} {\bibfnamefont {S.}~\bibnamefont {Ding}},
  \bibinfo {author} {\bibfnamefont {R.}~\bibnamefont {Hablutzel}}, \bibinfo
  {author} {\bibfnamefont {J.}~\bibnamefont {Gan}}, \bibinfo {author}
  {\bibfnamefont {A.}~\bibnamefont {Roulet}}, \bibinfo {author} {\bibfnamefont
  {S.}~\bibnamefont {Nimmrichter}}, \bibinfo {author} {\bibfnamefont
  {J.}~\bibnamefont {Dai}}, \bibinfo {author} {\bibfnamefont {V.}~\bibnamefont
  {Scarani}}, \ and\ \bibinfo {author} {\bibfnamefont {D.}~\bibnamefont
  {Matsukevich}},\ }\href {http://arxiv.org/abs/1702.08672} {\  (\bibinfo
  {year} {2017})},\ \Eprint {http://arxiv.org/abs/1702.08672}
  {arXiv:1702.08672} \BibitemShut {NoStop}%
\bibitem [{\citenamefont {Walls}\ and\ \citenamefont
  {Barakat}(1970)}]{Walls+Barakat:70}%
  \BibitemOpen
  \bibfield  {author} {\bibinfo {author} {\bibfnamefont {D.~F.}\ \bibnamefont
  {Walls}}\ and\ \bibinfo {author} {\bibfnamefont {R.}~\bibnamefont
  {Barakat}},\ }\href {\doibase 10.1103/PhysRevA.1.446} {\bibfield  {journal}
  {\bibinfo  {journal} {Phys. Rev. A}\ }\textbf {\bibinfo {volume} {1}},\
  \bibinfo {pages} {446} (\bibinfo {year} {1970})}\BibitemShut {NoStop}%
\bibitem [{\citenamefont {Agrawal}\ and\ \citenamefont
  {Mehta}()}]{Agrawal+Mehta:74}%
  \BibitemOpen
  \bibfield  {author} {\bibinfo {author} {\bibfnamefont {G.~P.}\ \bibnamefont
  {Agrawal}}\ and\ \bibinfo {author} {\bibfnamefont {C.~L.}\ \bibnamefont
  {Mehta}},\ }\href {\doibase 10.1088/0305-4470/7/5/011} {\bibinfo  {journal}
  {J. Phys. A}\ ,\ \bibinfo {pages} {607}}\BibitemShut {NoStop}%
\bibitem [{\citenamefont {Gambini}(1977)}]{Gambini:77}%
  \BibitemOpen
\bibfield  {journal} {  }\bibfield  {author} {\bibinfo {author} {\bibfnamefont
  {R.}~\bibnamefont {Gambini}},\ }\href {\doibase 10.1103/PhysRevA.15.1157}
  {\bibfield  {journal} {\bibinfo  {journal} {Phys. Rev. A}\ }\textbf {\bibinfo
  {volume} {15}},\ \bibinfo {pages} {1157} (\bibinfo {year}
  {1977})}\BibitemShut {NoStop}%
\bibitem [{\citenamefont {Levy}\ and\ \citenamefont
  {Kosloff}(2014)}]{Levy+Kosloff:14}%
  \BibitemOpen
  \bibfield  {author} {\bibinfo {author} {\bibfnamefont {A.}~\bibnamefont
  {Levy}}\ and\ \bibinfo {author} {\bibfnamefont {R.}~\bibnamefont {Kosloff}},\
  }\href {\doibase 10.1209/0295-5075/107/20004} {\bibfield  {journal} {\bibinfo
   {journal} {Europhys.~Lett.}\ }\textbf {\bibinfo {volume} {107}},\ \bibinfo
  {pages} {20004} (\bibinfo {year} {2014})}\BibitemShut {NoStop}%
\bibitem [{\citenamefont {Gonz{\'{a}}lez}\ \emph {et~al.}(2017)\citenamefont
  {Gonz{\'{a}}lez}, \citenamefont {Correa}, \citenamefont {Nocerino},
  \citenamefont {Palao}, \citenamefont {Alonso},\ and\ \citenamefont
  {Adesso}}]{Gonzalez2017}%
  \BibitemOpen
  \bibfield  {author} {\bibinfo {author} {\bibfnamefont {J.~O.}\ \bibnamefont
  {Gonz{\'{a}}lez}}, \bibinfo {author} {\bibfnamefont {L.~A.}\ \bibnamefont
  {Correa}}, \bibinfo {author} {\bibfnamefont {G.}~\bibnamefont {Nocerino}},
  \bibinfo {author} {\bibfnamefont {J.~P.}\ \bibnamefont {Palao}}, \bibinfo
  {author} {\bibfnamefont {D.}~\bibnamefont {Alonso}}, \ and\ \bibinfo {author}
  {\bibfnamefont {G.}~\bibnamefont {Adesso}},\ }\href
  {http://arxiv.org/abs/1707.09228} {\  (\bibinfo {year} {2017})},\ \Eprint
  {http://arxiv.org/abs/1707.09228} {arXiv:1707.09228} \BibitemShut {NoStop}%
\bibitem [{\citenamefont {Hofer}\ \emph {et~al.}(2017)\citenamefont {Hofer},
  \citenamefont {Perarnau-Llobet}, \citenamefont {Miranda}, \citenamefont
  {Haack}, \citenamefont {Silva}, \citenamefont {Brask},\ and\ \citenamefont
  {Brunner}}]{Hofer2017}%
  \BibitemOpen
  \bibfield  {author} {\bibinfo {author} {\bibfnamefont {P.~P.}\ \bibnamefont
  {Hofer}}, \bibinfo {author} {\bibfnamefont {M.}~\bibnamefont
  {Perarnau-Llobet}}, \bibinfo {author} {\bibfnamefont {L.~D.~M.}\ \bibnamefont
  {Miranda}}, \bibinfo {author} {\bibfnamefont {G.}~\bibnamefont {Haack}},
  \bibinfo {author} {\bibfnamefont {R.}~\bibnamefont {Silva}}, \bibinfo
  {author} {\bibfnamefont {J.~B.}\ \bibnamefont {Brask}}, \ and\ \bibinfo
  {author} {\bibfnamefont {N.}~\bibnamefont {Brunner}},\ }\href
  {http://arxiv.org/abs/1707.09211} {\  (\bibinfo {year} {2017})},\ \Eprint
  {http://arxiv.org/abs/1707.09211} {arXiv:1707.09211} \BibitemShut {NoStop}%
\bibitem [{\citenamefont {Bonifacio}\ and\ \citenamefont
  {Preparata}(1970)}]{Bonifacio+Preparata:70}%
  \BibitemOpen
  \bibfield  {author} {\bibinfo {author} {\bibfnamefont {R.}~\bibnamefont
  {Bonifacio}}\ and\ \bibinfo {author} {\bibfnamefont {G.}~\bibnamefont
  {Preparata}},\ }\href {\doibase 10.1103/PhysRevA.2.336} {\bibfield  {journal}
  {\bibinfo  {journal} {Phys. Rev. A}\ }\textbf {\bibinfo {volume} {2}},\
  \bibinfo {pages} {336} (\bibinfo {year} {1970})}\BibitemShut {NoStop}%
\bibitem [{\citenamefont {Gogolin}\ and\ \citenamefont
  {Eisert}(2016)}]{Gogolin+Eisert:16}%
  \BibitemOpen
  \bibfield  {author} {\bibinfo {author} {\bibfnamefont {C.}~\bibnamefont
  {Gogolin}}\ and\ \bibinfo {author} {\bibfnamefont {J.}~\bibnamefont
  {Eisert}},\ }\href {\doibase 10.1088/0034-4885/79/5/056001} {\bibfield
  {journal} {\bibinfo  {journal} {Rep. Prog. Phys.}\ }\textbf {\bibinfo
  {volume} {79}},\ \bibinfo {pages} {56001} (\bibinfo {year}
  {2016})}\BibitemShut {NoStop}%
\bibitem [{\citenamefont {Brunner}\ \emph {et~al.}(2012)\citenamefont
  {Brunner}, \citenamefont {Linden}, \citenamefont {Popescu},\ and\
  \citenamefont {Skrzypczyk}}]{Brunner+3:12}%
  \BibitemOpen
  \bibfield  {author} {\bibinfo {author} {\bibfnamefont {N.}~\bibnamefont
  {Brunner}}, \bibinfo {author} {\bibfnamefont {N.}~\bibnamefont {Linden}},
  \bibinfo {author} {\bibfnamefont {S.}~\bibnamefont {Popescu}}, \ and\
  \bibinfo {author} {\bibfnamefont {P.}~\bibnamefont {Skrzypczyk}},\ }\href
  {\doibase 10.1103/PhysRevE.85.051117} {\bibfield  {journal} {\bibinfo
  {journal} {Phys. Rev. E}\ }\textbf {\bibinfo {volume} {85}},\ \bibinfo
  {pages} {051117} (\bibinfo {year} {2012})}\BibitemShut {NoStop}%
\bibitem [{\citenamefont {Brask}\ and\ \citenamefont
  {Brunner}(2015)}]{Brask+Brunner:15}%
  \BibitemOpen
  \bibfield  {author} {\bibinfo {author} {\bibfnamefont {J.~B.}\ \bibnamefont
  {Brask}}\ and\ \bibinfo {author} {\bibfnamefont {N.}~\bibnamefont
  {Brunner}},\ }\href {\doibase 10.1103/PhysRevE.92.062101} {\bibfield
  {journal} {\bibinfo  {journal} {Phys. Rev. E}\ }\textbf {\bibinfo {volume}
  {92}},\ \bibinfo {pages} {062101} (\bibinfo {year} {2015})}\BibitemShut
  {NoStop}%
\bibitem [{\citenamefont {Mitchison}\ \emph {et~al.}(2015)\citenamefont
  {Mitchison}, \citenamefont {Woods}, \citenamefont {Prior},\ and\
  \citenamefont {Huber}}]{Mitchison+3:15}%
  \BibitemOpen
  \bibfield  {author} {\bibinfo {author} {\bibfnamefont {M.~T.}\ \bibnamefont
  {Mitchison}}, \bibinfo {author} {\bibfnamefont {M.~P.}\ \bibnamefont
  {Woods}}, \bibinfo {author} {\bibfnamefont {J.}~\bibnamefont {Prior}}, \ and\
  \bibinfo {author} {\bibfnamefont {M.}~\bibnamefont {Huber}},\ }\href
  {\doibase 10.1088/1367-2630/17/11/115013} {\bibfield  {journal} {\bibinfo
  {journal} {New J. Phys.}\ }\textbf {\bibinfo {volume} {17}},\ \bibinfo
  {pages} {115013} (\bibinfo {year} {2015})}\BibitemShut {NoStop}%
\bibitem [{\citenamefont {Kulkarni}\ \emph {et~al.}(2012)\citenamefont
  {Kulkarni}, \citenamefont {Tiwari},\ and\ \citenamefont
  {Segal}}]{Kulkarni+2:12}%
  \BibitemOpen
  \bibfield  {author} {\bibinfo {author} {\bibfnamefont {M.}~\bibnamefont
  {Kulkarni}}, \bibinfo {author} {\bibfnamefont {K.~L.}\ \bibnamefont
  {Tiwari}}, \ and\ \bibinfo {author} {\bibfnamefont {D.}~\bibnamefont
  {Segal}},\ }\href {\doibase 10.1103/PhysRevB.86.155424} {\bibfield  {journal}
  {\bibinfo  {journal} {Phys. Rev. B}\ }\textbf {\bibinfo {volume} {86}},\
  \bibinfo {pages} {155424} (\bibinfo {year} {2012})}\BibitemShut {NoStop}%
\bibitem [{\citenamefont {Farrelly}\ \emph {et~al.}(2017)\citenamefont
  {Farrelly}, \citenamefont {Brand{\~{a}}o},\ and\ \citenamefont
  {Cramer}}]{Farrelly+2:17}%
  \BibitemOpen
  \bibfield  {author} {\bibinfo {author} {\bibfnamefont {T.}~\bibnamefont
  {Farrelly}}, \bibinfo {author} {\bibfnamefont {F.~G. S.~L.}\ \bibnamefont
  {Brand{\~{a}}o}}, \ and\ \bibinfo {author} {\bibfnamefont {M.}~\bibnamefont
  {Cramer}},\ }\href {\doibase 10.1103/PhysRevLett.118.140601} {\bibfield
  {journal} {\bibinfo  {journal} {Phys. Rev. Lett.}\ }\textbf {\bibinfo
  {volume} {118}},\ \bibinfo {pages} {140601} (\bibinfo {year}
  {2017})}\BibitemShut {NoStop}%
\bibitem [{\citenamefont {Erez}\ \emph {et~al.}(2008)\citenamefont {Erez},
  \citenamefont {Gordon}, \citenamefont {Nest},\ and\ \citenamefont
  {Kurizki}}]{Erez2008}%
  \BibitemOpen
  \bibfield  {author} {\bibinfo {author} {\bibfnamefont {N.}~\bibnamefont
  {Erez}}, \bibinfo {author} {\bibfnamefont {G.}~\bibnamefont {Gordon}},
  \bibinfo {author} {\bibfnamefont {M.}~\bibnamefont {Nest}}, \ and\ \bibinfo
  {author} {\bibfnamefont {G.}~\bibnamefont {Kurizki}},\ }\href {\doibase
  10.1038/nature06873} {\bibfield  {journal} {\bibinfo  {journal} {Nature}\
  }\textbf {\bibinfo {volume} {452}},\ \bibinfo {pages} {724} (\bibinfo {year}
  {2008})}\BibitemShut {NoStop}%
\bibitem [{\citenamefont {Glauber}(2007)}]{Glauber2006}%
  \BibitemOpen
  \bibfield  {author} {\bibinfo {author} {\bibfnamefont {R.~J.}\ \bibnamefont
  {Glauber}},\ }\href {\doibase 10.1002/9783527610075} {\emph {\bibinfo {title}
  {{Quantum Theory of Optical Coherence}}}}\ (\bibinfo  {publisher}
  {Wiley-VCH},\ \bibinfo {year} {2007})\BibitemShut {NoStop}%
\bibitem [{\citenamefont {Jensen}\ and\ \citenamefont
  {Shankar}(1985)}]{Jensen+Shankar:85}%
  \BibitemOpen
  \bibfield  {author} {\bibinfo {author} {\bibfnamefont {R.~V.}\ \bibnamefont
  {Jensen}}\ and\ \bibinfo {author} {\bibfnamefont {R.}~\bibnamefont
  {Shankar}},\ }\href {\doibase 10.1103/PhysRevLett.54.1879} {\bibfield
  {journal} {\bibinfo  {journal} {Phys. Rev. Lett.}\ }\textbf {\bibinfo
  {volume} {54}},\ \bibinfo {pages} {1879} (\bibinfo {year}
  {1985})}\BibitemShut {NoStop}%
\bibitem [{\citenamefont {Tasaki}(1998)}]{Tasaki1998}%
  \BibitemOpen
  \bibfield  {author} {\bibinfo {author} {\bibfnamefont {H.}~\bibnamefont
  {Tasaki}},\ }\href {\doibase 10.1103/PhysRevLett.80.1373} {\bibfield
  {journal} {\bibinfo  {journal} {Phys. Rev. Lett.}\ }\textbf {\bibinfo
  {volume} {80}},\ \bibinfo {pages} {1373} (\bibinfo {year}
  {1998})}\BibitemShut {NoStop}%
\bibitem [{\citenamefont {Reimann}(2008)}]{Reimann:08}%
  \BibitemOpen
  \bibfield  {author} {\bibinfo {author} {\bibfnamefont {P.}~\bibnamefont
  {Reimann}},\ }\href {\doibase 10.1103/PhysRevLett.101.190403} {\bibfield
  {journal} {\bibinfo  {journal} {Phys. Rev. Lett.}\ }\textbf {\bibinfo
  {volume} {101}},\ \bibinfo {pages} {190403} (\bibinfo {year}
  {2008})}\BibitemShut {NoStop}%
\bibitem [{\citenamefont {Rigol}\ \emph {et~al.}(2008)\citenamefont {Rigol},
  \citenamefont {Dunjko},\ and\ \citenamefont {Olshanii}}]{Rigol+2:08}%
  \BibitemOpen
  \bibfield  {author} {\bibinfo {author} {\bibfnamefont {M.}~\bibnamefont
  {Rigol}}, \bibinfo {author} {\bibfnamefont {V.}~\bibnamefont {Dunjko}}, \
  and\ \bibinfo {author} {\bibfnamefont {M.}~\bibnamefont {Olshanii}},\ }\href
  {\doibase 10.1038/nature06838} {\bibfield  {journal} {\bibinfo  {journal}
  {Nature}\ }\textbf {\bibinfo {volume} {452}},\ \bibinfo {pages} {854}
  (\bibinfo {year} {2008})}\BibitemShut {NoStop}%
\bibitem [{\citenamefont {Short}(2011)}]{Short:11}%
  \BibitemOpen
  \bibfield  {author} {\bibinfo {author} {\bibfnamefont {A.~J.}\ \bibnamefont
  {Short}},\ }\href {\doibase 10.1088/1367-2630/13/5/053009} {\bibfield
  {journal} {\bibinfo  {journal} {New J. Phys.}\ }\textbf {\bibinfo {volume}
  {13}},\ \bibinfo {pages} {53009} (\bibinfo {year} {2011})}\BibitemShut
  {NoStop}%
\bibitem [{\citenamefont {Ponomarev}\ \emph {et~al.}(2011)\citenamefont
  {Ponomarev}, \citenamefont {Denisov},\ and\ \citenamefont
  {H{\"{a}}nggi}}]{Ponomarev+2:11}%
  \BibitemOpen
  \bibfield  {author} {\bibinfo {author} {\bibfnamefont {A.~V.}\ \bibnamefont
  {Ponomarev}}, \bibinfo {author} {\bibfnamefont {S.}~\bibnamefont {Denisov}},
  \ and\ \bibinfo {author} {\bibfnamefont {P.}~\bibnamefont {H{\"{a}}nggi}},\
  }\href {\doibase 10.1103/PhysRevLett.106.010405} {\bibfield  {journal}
  {\bibinfo  {journal} {Phys. Rev. Lett.}\ }\textbf {\bibinfo {volume} {106}},\
  \bibinfo {pages} {010405} (\bibinfo {year} {2011})}\BibitemShut {NoStop}%
\bibitem [{\citenamefont {Short}\ and\ \citenamefont
  {Farrelly}(2012)}]{Short+Farrelly:12}%
  \BibitemOpen
  \bibfield  {author} {\bibinfo {author} {\bibfnamefont {A.~J.}\ \bibnamefont
  {Short}}\ and\ \bibinfo {author} {\bibfnamefont {T.~C.}\ \bibnamefont
  {Farrelly}},\ }\href {\doibase 10.1088/1367-2630/14/1/013063} {\bibfield
  {journal} {\bibinfo  {journal} {New J. Phys.}\ }\textbf {\bibinfo {volume}
  {14}},\ \bibinfo {pages} {13063} (\bibinfo {year} {2012})}\BibitemShut
  {NoStop}%
\bibitem [{\citenamefont {Reimann}(2012)}]{Reimann:12}%
  \BibitemOpen
  \bibfield  {author} {\bibinfo {author} {\bibfnamefont {P.}~\bibnamefont
  {Reimann}},\ }\href {\doibase 10.1088/0031-8949/86/05/058512} {\bibfield
  {journal} {\bibinfo  {journal} {Phys. Scr.}\ }\textbf {\bibinfo {volume}
  {86}},\ \bibinfo {pages} {58512} (\bibinfo {year} {2012})}\BibitemShut
  {NoStop}%
\bibitem [{\citenamefont {Malabarba}\ \emph {et~al.}(2014)\citenamefont
  {Malabarba}, \citenamefont {Garc{\'{i}}a-Pintos}, \citenamefont {Linden},
  \citenamefont {Farrelly},\ and\ \citenamefont {Short}}]{Malabarba+4:14}%
  \BibitemOpen
  \bibfield  {author} {\bibinfo {author} {\bibfnamefont {A.~S.~L.}\
  \bibnamefont {Malabarba}}, \bibinfo {author} {\bibfnamefont {L.~P.}\
  \bibnamefont {Garc{\'{i}}a-Pintos}}, \bibinfo {author} {\bibfnamefont
  {N.}~\bibnamefont {Linden}}, \bibinfo {author} {\bibfnamefont {T.~C.}\
  \bibnamefont {Farrelly}}, \ and\ \bibinfo {author} {\bibfnamefont {A.~J.}\
  \bibnamefont {Short}},\ }\href {\doibase 10.1103/PhysRevE.90.012121}
  {\bibfield  {journal} {\bibinfo  {journal} {Phys. Rev. E}\ }\textbf {\bibinfo
  {volume} {90}},\ \bibinfo {pages} {012121} (\bibinfo {year}
  {2014})}\BibitemShut {NoStop}%
\bibitem [{\citenamefont {Malabarba}\ \emph {et~al.}(2015)\citenamefont
  {Malabarba}, \citenamefont {Linden},\ and\ \citenamefont
  {Short}}]{Malabarba+2:15}%
  \BibitemOpen
  \bibfield  {author} {\bibinfo {author} {\bibfnamefont {A.~S.~L.}\
  \bibnamefont {Malabarba}}, \bibinfo {author} {\bibfnamefont {N.}~\bibnamefont
  {Linden}}, \ and\ \bibinfo {author} {\bibfnamefont {A.~J.}\ \bibnamefont
  {Short}},\ }\href {\doibase 10.1103/PhysRevE.92.062128} {\bibfield  {journal}
  {\bibinfo  {journal} {Phys. Rev. E}\ }\textbf {\bibinfo {volume} {92}},\
  \bibinfo {pages} {062128} (\bibinfo {year} {2015})}\BibitemShut {NoStop}%
\bibitem [{\citenamefont {Goldstein}\ \emph {et~al.}(2015)\citenamefont
  {Goldstein}, \citenamefont {Hara},\ and\ \citenamefont
  {Tasaki}}]{Goldstein+2:15}%
  \BibitemOpen
  \bibfield  {author} {\bibinfo {author} {\bibfnamefont {S.}~\bibnamefont
  {Goldstein}}, \bibinfo {author} {\bibfnamefont {T.}~\bibnamefont {Hara}}, \
  and\ \bibinfo {author} {\bibfnamefont {H.}~\bibnamefont {Tasaki}},\ }\href
  {\doibase 10.1088/1367-2630/17/4/045002} {\bibfield  {journal} {\bibinfo
  {journal} {New J. Phys.}\ }\textbf {\bibinfo {volume} {17}},\ \bibinfo
  {pages} {45002} (\bibinfo {year} {2015})}\BibitemShut {NoStop}%
\bibitem [{\citenamefont {Kaufman}\ \emph {et~al.}(2016)\citenamefont
  {Kaufman}, \citenamefont {Tai}, \citenamefont {Lukin}, \citenamefont
  {Rispoli}, \citenamefont {Schittko}, \citenamefont {Preiss},\ and\
  \citenamefont {Greiner}}]{Kaufman+6:16}%
  \BibitemOpen
  \bibfield  {author} {\bibinfo {author} {\bibfnamefont {A.~M.}\ \bibnamefont
  {Kaufman}}, \bibinfo {author} {\bibfnamefont {M.~E.}\ \bibnamefont {Tai}},
  \bibinfo {author} {\bibfnamefont {A.}~\bibnamefont {Lukin}}, \bibinfo
  {author} {\bibfnamefont {M.}~\bibnamefont {Rispoli}}, \bibinfo {author}
  {\bibfnamefont {R.}~\bibnamefont {Schittko}}, \bibinfo {author}
  {\bibfnamefont {P.~M.}\ \bibnamefont {Preiss}}, \ and\ \bibinfo {author}
  {\bibfnamefont {M.}~\bibnamefont {Greiner}},\ }\href {\doibase
  10.1126/science.aaf6725} {\bibfield  {journal} {\bibinfo  {journal}
  {Science}\ }\textbf {\bibinfo {volume} {353}},\ \bibinfo {pages} {794}
  (\bibinfo {year} {2016})}\BibitemShut {NoStop}%
\bibitem [{\citenamefont {Garc{\'{i}}a-Pintos}\ \emph
  {et~al.}(2017)\citenamefont {Garc{\'{i}}a-Pintos}, \citenamefont {Linden},
  \citenamefont {Malabarba}, \citenamefont {Short},\ and\ \citenamefont
  {Winter}}]{Pintos+4:15}%
  \BibitemOpen
  \bibfield  {author} {\bibinfo {author} {\bibfnamefont {L.~P.}\ \bibnamefont
  {Garc{\'{i}}a-Pintos}}, \bibinfo {author} {\bibfnamefont {N.}~\bibnamefont
  {Linden}}, \bibinfo {author} {\bibfnamefont {A.~S.~L.}\ \bibnamefont
  {Malabarba}}, \bibinfo {author} {\bibfnamefont {A.~J.}\ \bibnamefont
  {Short}}, \ and\ \bibinfo {author} {\bibfnamefont {A.}~\bibnamefont
  {Winter}},\ }\href {\doibase 10.1103/PhysRevX.7.031027} {\bibfield  {journal}
  {\bibinfo  {journal} {Phys. Rev. X}\ }\textbf {\bibinfo {volume} {7}},\
  \bibinfo {pages} {031027} (\bibinfo {year} {2017})}\BibitemShut {NoStop}%
\bibitem [{\citenamefont {Linden}\ \emph {et~al.}(2009)\citenamefont {Linden},
  \citenamefont {Popescu}, \citenamefont {Short},\ and\ \citenamefont
  {Winter}}]{Linden2009}%
  \BibitemOpen
  \bibfield  {author} {\bibinfo {author} {\bibfnamefont {N.}~\bibnamefont
  {Linden}}, \bibinfo {author} {\bibfnamefont {S.}~\bibnamefont {Popescu}},
  \bibinfo {author} {\bibfnamefont {A.~J.}\ \bibnamefont {Short}}, \ and\
  \bibinfo {author} {\bibfnamefont {A.}~\bibnamefont {Winter}},\ }\href
  {\doibase 10.1103/PhysRevE.79.061103} {\bibfield  {journal} {\bibinfo
  {journal} {Phys. Rev. E}\ }\textbf {\bibinfo {volume} {79}},\ \bibinfo
  {pages} {061103} (\bibinfo {year} {2009})}\BibitemShut {NoStop}%
\bibitem [{\citenamefont {Reimann}\ and\ \citenamefont
  {Kastner}(2012)}]{Reimann2012a}%
  \BibitemOpen
  \bibfield  {author} {\bibinfo {author} {\bibfnamefont {P.}~\bibnamefont
  {Reimann}}\ and\ \bibinfo {author} {\bibfnamefont {M.}~\bibnamefont
  {Kastner}},\ }\href {\doibase 10.1088/1367-2630/14/4/043020} {\bibfield
  {journal} {\bibinfo  {journal} {New J. Phys.}\ }\textbf {\bibinfo {volume}
  {14}},\ \bibinfo {pages} {43020} (\bibinfo {year} {2012})}\BibitemShut
  {NoStop}%
\bibitem [{\citenamefont {Shaffer}\ \emph {et~al.}(2014)\citenamefont
  {Shaffer}, \citenamefont {Chamon}, \citenamefont {Hamma},\ and\ \citenamefont
  {Mucciolo}}]{Shaffer+3:14}%
  \BibitemOpen
  \bibfield  {author} {\bibinfo {author} {\bibfnamefont {D.}~\bibnamefont
  {Shaffer}}, \bibinfo {author} {\bibfnamefont {C.}~\bibnamefont {Chamon}},
  \bibinfo {author} {\bibfnamefont {A.}~\bibnamefont {Hamma}}, \ and\ \bibinfo
  {author} {\bibfnamefont {E.~R.}\ \bibnamefont {Mucciolo}},\ }\href {\doibase
  10.1088/1742-5468/2014/12/P12007} {\bibfield  {journal} {\bibinfo  {journal}
  {J. Stat. Mech.}\ }\textbf {\bibinfo {volume} {2014}},\ \bibinfo {pages}
  {P12007} (\bibinfo {year} {2014})}\BibitemShut {NoStop}%
\bibitem [{\citenamefont {Sengupta}\ \emph {et~al.}(2004)\citenamefont
  {Sengupta}, \citenamefont {Powell},\ and\ \citenamefont
  {Sachdev}}]{Sengupta2004}%
  \BibitemOpen
  \bibfield  {author} {\bibinfo {author} {\bibfnamefont {K.}~\bibnamefont
  {Sengupta}}, \bibinfo {author} {\bibfnamefont {S.}~\bibnamefont {Powell}}, \
  and\ \bibinfo {author} {\bibfnamefont {S.}~\bibnamefont {Sachdev}},\ }\href
  {\doibase 10.1103/PhysRevA.69.053616} {\bibfield  {journal} {\bibinfo
  {journal} {Phys. Rev. A}\ }\textbf {\bibinfo {volume} {69}},\ \bibinfo
  {pages} {053616} (\bibinfo {year} {2004})}\BibitemShut {NoStop}%
\bibitem [{\citenamefont {Manmana}\ \emph {et~al.}(2007)\citenamefont
  {Manmana}, \citenamefont {Wessel}, \citenamefont {Noack},\ and\ \citenamefont
  {Muramatsu}}]{Manmana2007}%
  \BibitemOpen
  \bibfield  {author} {\bibinfo {author} {\bibfnamefont {S.~R.}\ \bibnamefont
  {Manmana}}, \bibinfo {author} {\bibfnamefont {S.}~\bibnamefont {Wessel}},
  \bibinfo {author} {\bibfnamefont {R.~M.}\ \bibnamefont {Noack}}, \ and\
  \bibinfo {author} {\bibfnamefont {A.}~\bibnamefont {Muramatsu}},\ }\href
  {\doibase 10.1103/PhysRevLett.98.210405} {\bibfield  {journal} {\bibinfo
  {journal} {Phys. Rev. Lett.}\ }\textbf {\bibinfo {volume} {98}},\ \bibinfo
  {pages} {210405} (\bibinfo {year} {2007})}\BibitemShut {NoStop}%
\bibitem [{\citenamefont {Luitz}\ \emph {et~al.}(2015)\citenamefont {Luitz},
  \citenamefont {Laflorencie},\ and\ \citenamefont {Alet}}]{Luitz2015}%
  \BibitemOpen
  \bibfield  {author} {\bibinfo {author} {\bibfnamefont {D.~J.}\ \bibnamefont
  {Luitz}}, \bibinfo {author} {\bibfnamefont {N.}~\bibnamefont {Laflorencie}},
  \ and\ \bibinfo {author} {\bibfnamefont {F.}~\bibnamefont {Alet}},\ }\href
  {\doibase 10.1103/PhysRevB.91.081103} {\bibfield  {journal} {\bibinfo
  {journal} {Phys. Rev. B}\ }\textbf {\bibinfo {volume} {91}},\ \bibinfo
  {pages} {081103} (\bibinfo {year} {2015})}\BibitemShut {NoStop}%
\bibitem [{\citenamefont {Kollath}\ \emph {et~al.}(2007)\citenamefont
  {Kollath}, \citenamefont {L{\"{a}}uchli},\ and\ \citenamefont
  {Altman}}]{Kollath2007}%
  \BibitemOpen
  \bibfield  {author} {\bibinfo {author} {\bibfnamefont {C.}~\bibnamefont
  {Kollath}}, \bibinfo {author} {\bibfnamefont {A.~M.}\ \bibnamefont
  {L{\"{a}}uchli}}, \ and\ \bibinfo {author} {\bibfnamefont {E.}~\bibnamefont
  {Altman}},\ }\href {\doibase 10.1103/PhysRevLett.98.180601} {\bibfield
  {journal} {\bibinfo  {journal} {Phys. Rev. Lett.}\ }\textbf {\bibinfo
  {volume} {98}},\ \bibinfo {pages} {180601} (\bibinfo {year}
  {2007})}\BibitemShut {NoStop}%
\bibitem [{\citenamefont {Rigol}(2009)}]{Rigol2009}%
  \BibitemOpen
  \bibfield  {author} {\bibinfo {author} {\bibfnamefont {M.}~\bibnamefont
  {Rigol}},\ }\href {\doibase 10.1103/PhysRevLett.103.100403} {\bibfield
  {journal} {\bibinfo  {journal} {Phys. Rev. Lett.}\ }\textbf {\bibinfo
  {volume} {103}},\ \bibinfo {pages} {100403} (\bibinfo {year}
  {2009})}\BibitemShut {NoStop}%
\bibitem [{\citenamefont {Ba{\~{n}}uls}\ \emph {et~al.}(2011)\citenamefont
  {Ba{\~{n}}uls}, \citenamefont {Cirac},\ and\ \citenamefont
  {Hastings}}]{Banuls2011}%
  \BibitemOpen
  \bibfield  {author} {\bibinfo {author} {\bibfnamefont {M.~C.}\ \bibnamefont
  {Ba{\~{n}}uls}}, \bibinfo {author} {\bibfnamefont {J.~I.}\ \bibnamefont
  {Cirac}}, \ and\ \bibinfo {author} {\bibfnamefont {M.~B.}\ \bibnamefont
  {Hastings}},\ }\href {\doibase 10.1103/PhysRevLett.106.050405} {\bibfield
  {journal} {\bibinfo  {journal} {Phys. Rev. Lett.}\ }\textbf {\bibinfo
  {volume} {106}},\ \bibinfo {pages} {050405} (\bibinfo {year}
  {2011})}\BibitemShut {NoStop}%
\bibitem [{\citenamefont {Bajer}\ and\ \citenamefont
  {Miranowicz}(2001)}]{Bajer+Miranowicz:01}%
  \BibitemOpen
  \bibfield  {author} {\bibinfo {author} {\bibfnamefont {J.}~\bibnamefont
  {Bajer}}\ and\ \bibinfo {author} {\bibfnamefont {A.}~\bibnamefont
  {Miranowicz}},\ }\href {\doibase 10.1088/1464-4266/3/4/309} {\bibfield
  {journal} {\bibinfo  {journal} {J. Opt. B Quantum Semiclass. Opt.}\ }\textbf
  {\bibinfo {volume} {3}},\ \bibinfo {pages} {251} (\bibinfo {year}
  {2001})}\BibitemShut {NoStop}%
\bibitem [{\citenamefont {Abramowitz}\ and\ \citenamefont
  {Stegun}(1965)}]{Abramowitz1965}%
  \BibitemOpen
  \bibfield  {author} {\bibinfo {author} {\bibfnamefont {M.}~\bibnamefont
  {Abramowitz}}\ and\ \bibinfo {author} {\bibfnamefont {I.~A.}\ \bibnamefont
  {Stegun}},\ }\href
  {https://www.worldcat.org/title/handbook-of-mathematical-functions-with-formulas-graphs-and-mathematical-tables/oclc/18003605}
  {\emph {\bibinfo {title} {{Handbook of Mathematical Functions: with Formulas,
  Graphs, and Mathematical Tables}}}}\ (\bibinfo  {publisher} {Dover, New
  York},\ \bibinfo {year} {1965})\BibitemShut {NoStop}%
\end{thebibliography}

%

\end{document}